\documentclass[11pt]{elsarticle}
\usepackage{bm}
\usepackage{graphicx}
\usepackage{amsmath}
\usepackage{cases}
\usepackage{amssymb}
\usepackage{booktabs}
\usepackage{multirow}
\usepackage{xcolor}
\usepackage{epstopdf}
\usepackage[normalem]{ulem}

\usepackage[caption=false]{subfig}
\usepackage[top=1.1in, bottom=1.1in, left=1.0in, right=1.0in]{geometry}
\usepackage{tikz}
\usepackage{comment}
\usetikzlibrary{calc}

\begin{document}
\title{Performance evaluation of the general characteristics based off-lattice Boltzmann and DUGKS methods for low speed continuum flows: A comparative study}
\author[]{Lianhua Zhu}
\ead{lhzhu@hust.edu.cn}
\author[]{Peng Wang}
\ead{sklccwangpeng@hust.edu.cn}
\author[]{Zhaoli Guo\corref{cor1}}
\ead{zlguo@hust.edu.cn}

\cortext[cor1]{Corresponding author}
  \address{State Key Laboratory of Coal Combustion, Huazhong University of Science and Technology, Wuhan, 430074, China}
\begin{abstract}
The general characteristics based off-lattice Boltzmann scheme (BKG) proposed by Bardow et~al.,~\cite{bardow06} and the discrete unified gas kinetic scheme (DUGKS)~\cite{guozl13}  are two methods  that successfully overcome the time step restriction by the collision time, which is commonly seen in  many other kinetic schemes.
Basically, the BKG scheme is a time splitting scheme, while the DUGKS is an un-split finite volume scheme.
In this work, we first perform a theoretical analysis of the two schemes in the finite volume framework by comparing their numerical flux evaluations.
It is found that the effects of collision term are considered in the reconstructions of the cell-interface distribution function in both schemes,
which explains why they can overcome the time step restriction and can give accurate results even as the time step is much larger than the collision time.
The difference between the two schemes lies in the treatment of the integral of the collision term,
in which the Bardow's scheme uses the rectangular rule while the DUGKS uses the trapezoidal rule.
The performance of the two schemes, i.e., accuracy, stability, and efficiency are then compared by simulating several two dimensional flows,
including the unsteady Taylor-Green vortex flow, the steady lid-driven cavity flow, and the laminar boundary layer problem.
It is observed that, the DUGKS can give more accurate results than the BKG scheme.
Furthermore, the numerical stability of the BKG scheme decreases as the Courant-Friedrichs-Lewy (CFL) number approaches to 1,
while the stability of DUGKS is not affected by the CFL number apparently as long as $\text{CFL}<1$.
It is also observed that the BKG scheme is about one time faster than the DUGKS scheme with the same computational mesh and time step.
\end{abstract}


\maketitle
\section{Introduction}
The lattice Boltzmann method (LBM) has become a popular numerical tool for flow simulations. It solves the discrete velocity Boltzmann equation (DVBE) with sophistically chosen discrete velocity set. With the coupled discretization of velocity space and spatial space, the numerical treatment of convection term reduces to a very simple \emph{streaming} processes, which provides the benefits of low numerical dissipation, easy implementation, and high parallel computing efficiency. Another advantage of LBM is that, the simplified collision term is computed implicitly while implemented explicitly, which allows for a large time step even though the collision term causess stiffness at a small relaxation time. This advantage makes the LBM a potential solver for high Reynolds number flows.

However, the coupled discretization of velocity and spatial spaces limits the LBM to the use of uniform Cartesian meshes, which prohibits its applications for practical engineering problems. Some efforts have been made to extend the standard LBM to non-regular (non-uniform, unstructured) meshes, and a number of so called off-lattice Boltzmann (OLB) methods have been developed by solving the DVBE using certain finite-difference, finite-volume, or finite-element schemes~\cite{bardow06, meirw98, penggw99, lee01, lee03, guozl03, rossi05, bardow08, patil09}. These OLB schemes differ from each other in the temporal and spatial discretizations. However, a straightforward implementation of the CFD techniques usually leads to the loss of the advantages of standard LBM, especially the low dissipation property and stability at large time step. For example, in many of the schemes~\cite{penggw99, guozl03, rossi05, patil09, patil12}, the time step is limited by the relaxation time to get an accurate solution, even as the collision term is computed implicitly~\cite{guozl03}. This drawback makes these OLB schemes very computational expensive when simulating high Reynolds number flows.

An alternative way to construct OLB schemes is to use the time-splitting strategy in solving the DVBE~\cite{bardow06,lee01,lee03,bardow08,mcnamara95,guozl01}, in which the DVBE is decomposed into a collision sub-equation and a followed pure advection sub-equation. The collision sub-equation is fully local and is discretized directly, leading to a collision step the same as the standard LBM; The collisionless advection subequation is then solved with certain numerical schemes on uniform or non-uniform meshes~\cite{lee01,lee03}, leading to a general streaming step. Specifically, the scheme proposed by Bardow et~al.~(denoted by BKG), which combines the variable transformation technique for the collision term and the Lax-Wendroff scheme for the streaming step, overcomes the time step restriction by the relaxation time. It was demonstrated that accurate and stable solutions can be obtained even as the time step is much larger than the relaxation time~\cite{bardow06,bardow08,rao15}.

The above OLB schemes are developed in the LBM framework, and are limited to continuum flows. Recently, a finite volume kinetic approach using general mesh, i.e., discrete unified gas kinetic scheme (DUGKS), was proposed for all Knudsen number flows~\cite{guozl13, guozl15}. In the DUGKS the numerical flux is constructed based on the governing equation i.e., the DVBE itself, instead of using interpolations. With such a treatment, the time step is not restricted by the relaxation time, and its superior accuracy and stability for high Reynolds continuum flows have been demonstrated~\cite{wangp15,zhulh15}.

Since both the BKG and the DUGKS methods overcome the time step restriction from different approaches, it is still not clear the performance difference between them,
so in this work we will present a comparative study of these two kinetic schemes for continuum flows, even the DUGKS is not limited to such flows. We will also investigate the link between the two schemes by comparing them in the same finite volume framework.

The remaining part of this paper is organized as follows. Sec.~2 will introduce the DUGKS and BKG methods and discuss their relation, Sec.~3 will present the comparison results, and a conclusion is given in Sec.~4.

\section{Numerical formulation}
\subsection{Discrete Velocity Boltzmann-BGK equation}
The governing equation for the OLB schemes and DUGKS method is the Boltzmann equation with the Bhatnagar-Gross-Krook collision operator~\cite{bgk},
\begin{equation}\label{BGK}
 \frac{\partial f}{\partial t}+{\bm \xi}\cdot \bm \nabla f=\Omega(f)\equiv\frac{f^{eq}-f}{\tau},
\end{equation}
where $f\equiv f(\bm{x},\bm{\xi},t)$ is the distribution function (DF) with particle velocity $\bm{\xi}$ at position $\bm{x}$ and time $t$,
$\tau$ is relaxation time due to particle collisions, and $f^{eq}$ is the Maxwellian equilibrium distribution function.
In this article, we consider the isothermal two-dimensional-nine-velocities (D2Q9) lattice model.
The corresponding DVBE is
\begin{equation}\label{BGK_discrete}
 \frac{\partial f_\alpha}{\partial t}+{\bm \xi_\alpha}\cdot\bm \nabla f_\alpha=\Omega(f_\alpha)\equiv\frac{f_\alpha^{eq}-f_\alpha}{\tau},
\end{equation}
where $f_\alpha \equiv f(\bm{x},\bm{\xi}_\alpha, t)$ and $f_\alpha^{eq} \equiv f^{eq}(\bm{x}, \bm{\xi}_\alpha, t)$ are the DF with discrete velocity $\bm{\xi}_\alpha$ and the
corresponding discrete equilibrium DF respectively.
The D2Q9 discrete velocity set $\bm{\xi}_\alpha$ is defined as
\begin{equation}
\bm{\xi}_\alpha=
\begin{cases}
  (0,0)                                            & \text{for} ~\alpha=0,\\
  \sqrt{3RT} \left(\cos[(\alpha-1)\pi/2], \sin[(\alpha-1)\pi/2]\right)            & \text{for} ~ \alpha=1,2,3,4,\\
  \sqrt{3RT} \left(\cos[(2\alpha-9)\pi/4], \sin[(2\alpha-9)\pi/4]\right)\sqrt{2}  & \text{for} ~ \alpha=5,6,7,8,
\end{cases}
\end{equation}
where $R$ is the gas constant and $T$ is the constant temperature.
Under the low Mach number condition, the discrete equilibrium DF
can be approximated by its Taylor expansion around zero particle velocity up to second order,
\begin{equation}\label{equilibrium}
f^{eq}_\alpha=w_\alpha\rho\left[1+\frac{\bm{\xi}_\alpha\cdot\bm{u}}{c_s^2}+ \frac{(\bm{\xi}_\alpha \cdot\bm{u})^2}{2c_s^4}-\frac{\mid\bm{u}\mid^2}{2c_s^2} \right],
\end{equation}
where $c_s = \sqrt{RT}$ is the lattice sound speed and the weights $w_\alpha$ are
\begin{equation}
w_\alpha=
\begin{cases}
  4/9  & \text{for} ~ \alpha=0,\\
  1/9  & \text{for} ~ \alpha=1,2,3,4,\\
  1/36 & \text{for} ~ \alpha=5,6,7,8.
\end{cases}
\end{equation}
The fluid density $\rho$ and velocity $\bm u$ are the moments of the DF,
\begin{equation}\label{eq:macro}
  \rho = \sum_\alpha f_\alpha, \quad \rho\bm u = \sum_\alpha\bm \xi_\alpha f_\alpha.
\end{equation}
The shear viscosity of the fluid is related to the relaxation time by
\begin{equation}
  \nu = \tau RT,
  \label{nu_tau}
\end{equation}
which can be deduced from Chapman-Enskog analysis~\cite{guobook}.
The conservation property of the collision term is maintained at its discrete velocity counterpart, i.e.,
\begin{equation}\label{eq:convervation}
  \sum_\alpha \Omega(f_\alpha) = 0, \quad \sum_\alpha \bm \xi_\alpha \Omega(f_\alpha) = 0.
\end{equation}

\subsection{Discrete unified gas kinetic scheme}
The DUGKS employs a cell centered finite volume (FV) discretization of the DVBE~\cite{guozl13}. The computational domain is firstly divided into small control volumes $V_k$.
For a clear illustration of the formulas, we denote the volume averaged DF with discrete velocity $\xi_\alpha$ in control volume $V_k$ at time level $t^n$ by $f^n_{\alpha,k}$, i.e.,
\begin{equation}\label{eq:dugks_fv}
f^n_{\alpha,k} = \frac{1}{|V_k|}\int_{V_k} f_\alpha(\bm x, t^n) dV.
\end{equation}
Then integrating Eq.~\eqref{BGK_discrete} from time $t^n$ to time $t^{n+1}$ and applying the Gauss theorem we can get
\begin{equation}\label{eq_raw}
  f^{n+1}_{\alpha,k} - f^{n}_{\alpha,k} = \frac{\Delta t}{|V_k|}\mathcal{F}_{\alpha,k,\text{dugks}} + \frac{\Delta t}{2}\left[ \Omega(f^n_{\alpha,k}) + \Omega(f^{n+1}_{\alpha,k})\right ],
\end{equation}
where $\mathcal{F}_{\alpha,k,\text{dugks}}$ is the numerical flux that flows into the control volume from its faces, and $\Delta t= t^{n+1} - t^n$ is the time step size.
Note that trapezoidal rule is used for the collision term.
This implicit treatment of the collision term is crucial for its stability when the time step is much larger than the relaxation time.
This implicitness can be removed in the actual implementation using the following variable transformation technique, which is also adopted by the standard LBM,
\begin{equation}
\tilde f = f - \Delta t/2 \Omega(f), \quad \tilde f^+ = f + \Delta t/2 \Omega(f).
\end{equation}
Equation~\eqref{eq_raw} can then be rewritten in an explicit formulation,
\begin{equation}\label{eq_update}
  \tilde f^{n+1}_{\alpha,k} = \tilde f^{n,+}_{\alpha,k} + \frac{\Delta t}{|V_k|}\mathcal{F}_{\alpha,k,\text{dugks}}.
\end{equation}
In the implementation, we track the evolution of $\tilde f_{\alpha,k}$ instead of the original DF.
Due to the conservation property of the collision term, the macroscopic variables can be calculated by the transformed DF,
\begin{equation}\label{eq:macro_var}
  \rho = \sum_\alpha \tilde f_\alpha, \quad \rho\bm u = \sum_\alpha\bm \xi_\alpha \tilde f_\alpha.
\end{equation}

The key merit of DUGKS lies in its treatment of the advection term, i.e., the way to construct the numerical flux $\mathcal{F}_{\alpha,k,\text{dugks}}$.
In DUGKS, the middle point rule is used for the integration of the flux,
\begin{equation}
  \mathcal{F}_{\alpha,k,\text{dugks}} = \int_{\partial V_k} ( \bm \xi_\alpha \cdot \bm n)f^{n+1/2}_{\alpha,\text{dugks}}dS.
  \label{eq_flux}
\end{equation}
The integration over the faces is computed by the $f^{n+1/2}_\alpha$ at the centers of the faces,
which are computed using the characteristic solution of the kinetic equation~\eqref{eq_raw}.
Supposed the center of a face is $\bm x_b$,
then integrating Eq.~\eqref{eq_raw} along the characteristic line in a half time step $h=\Delta t/2$ from $(t^n, \bm x_b - h\bm \xi_\alpha )$ to $(t^{n+1/2}, \bm x_b)$,
and applying the trapezoidal rule, we can get
\begin{equation}
  f^{n+1/2}_{\alpha,\text{dugks}}(\bm x_b ) - f^n_\alpha(\bm x_b- h\bm\xi_\alpha) = \frac{h}{2}\left[ \Omega (f^n_\alpha(\bm x_b) ) + \Omega \left(f^{n+1/2}_\alpha(\bm x_b - h \bm \xi_\alpha ) \right)  \right ].
  \label{eq_charraw}
\end{equation}
Again the implicitness can be eliminated by introducing another two variable transformations,
\begin{equation}
  \bar f = f - h/2 \Omega(f), \quad \bar f^+ = f + h/2 \Omega(f),
  \label{eq_trans_2}
\end{equation}
and we can reformulate Eq.~\eqref{eq_charraw} in an explicit form,
\begin{equation}
  \bar f^{n+1/2}_\alpha(\bm x_b ) = \bar f^{n,+}_\alpha(\bm x_b- h\bm\xi_\alpha).
  \label{eq_char}
\end{equation}
For smooth flow, $\bar f^{n,+}_\alpha(\bm x_b- h\bm\xi)$ can be interpolated linearly from its neighboring cell centers.
After getting $\bar f^{n+1/2}_\alpha(\bm x_b )$,  the original DF $f^{n+1/2}_\alpha(\bm x_b)$ can be transformed back with the help of Eq.~\eqref{eq_trans_2}.
The macroscopic fluid variable $\rho^{n+1/2}(\bm x_b)$ and $\bm u^{n+1/2}(\bm x_b)$ used by the collision term in Eq.~\eqref{eq_trans_2} are calculated from,
\begin{equation}
 \left. \rho^{n+1/2}\right|_{\bm x_b} = \sum_\alpha \bar f_\alpha^{n+1/2}(\bm x_b), \quad
 \left.(\rho\bm u)^{n+1/2}\right|_{\bm x_b} = \sum_\alpha\bm \xi_\alpha \bar f_\alpha^{n+1/2}(\bm x_b).
  \label{}
\end{equation}
To insure the interpolation is stable, the time step is limited by the CFL condition,
\begin{equation}
  \Delta t= \eta \frac{\Delta x}{|\bm \xi|_{\text{max}}} = \eta \frac{\Delta x}{\sqrt{6}c_s},
  \label{eq:cfl}
\end{equation}
where $0<\eta <1$ is the CFL number and $\Delta x$ measures the size of the cell.

\subsection{The BKG scheme}
\begin{figure}[htbp]
\centering
\includegraphics[width=0.4\textwidth]{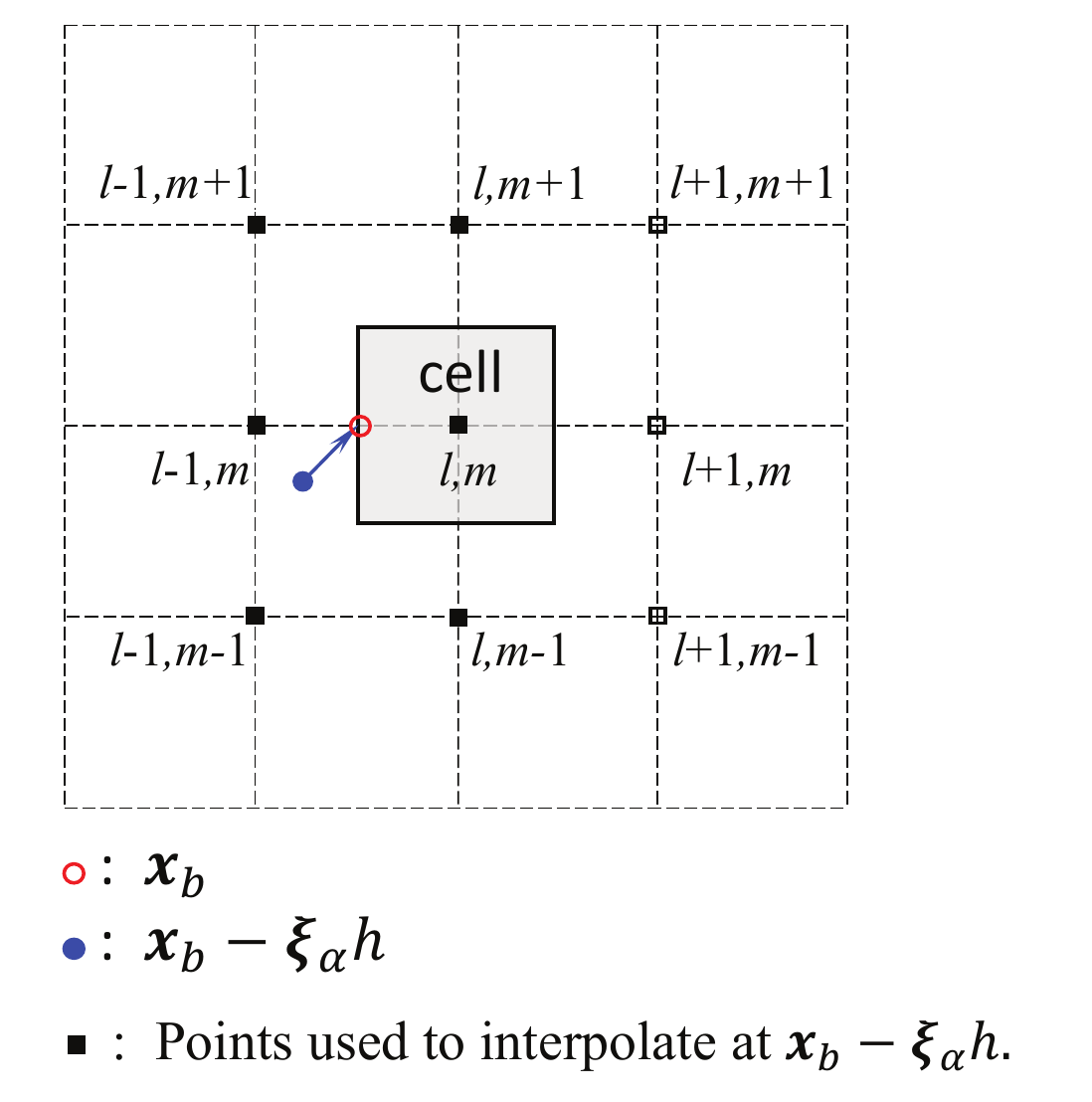}
\caption{
  Illustration of the finite volume discretization and the interpolation scheme.
} \label{fig:fv_stencil}
\end{figure}

Unlike the DUGKS in which the collision and particle-transport are treated simultaneously,
the BKG scheme is a splitting method for Eq.~\eqref{BGK_discrete}, which treats the convection term and collision term sequentially, i.e.,
\begin{subequations}
\begin{align}\label{eq:split_a}
 \frac{\partial f_\alpha}{\partial t}=\Omega(f_\alpha),\\
 \label{eq:split_b}
 \frac{\partial \tilde f_\alpha}{\partial t}+{\bm \xi_\alpha}\cdot\bm \nabla \tilde f_\alpha=0.
\end{align}
\end{subequations}
In the collision step, the collision term is integrated using trapezoidal rule~\cite{bardow06},
  \begin{equation}
    f^{*}_\alpha- f_\alpha^n = \frac{\Delta t}{2}\left [ \Omega(f^*_\alpha) + \Omega(f^n_\alpha) \right ].
    \label{eq:splic_c}
  \end{equation}
Using the same notation as used in the DUGKS, Eq.~\eqref{eq:splic_c} can be rewritten in an explicit formulation,
\begin{equation}
  \tilde f^{*}_\alpha = \tilde f^{+,n}_\alpha,
  \label{eq:bkg_collision}
\end{equation}
with $\tilde f^*_\alpha\equiv f^*_\alpha - (\Delta t/2)\Omega(f^*_\alpha) $.
It is noted that this treatment is identical to that of the standard LBM.
Then Eq.~\eqref{eq:split_b} is solved with the Lax-Wendroff scheme with the initial value $\tilde f^*$ or $\tilde f^{+,n}$,
\begin{equation}
\tilde f^{n+1}_\alpha = \tilde f^{+,n}_\alpha - \Delta t  \xi_{\alpha i} \frac{\partial \tilde f^{+,n}_\alpha}{\partial x_i}
+ \frac{\Delta t^2}{2}\xi_{\alpha i}\xi_{\alpha j} \frac{\partial^2 \tilde f^{+,n}_\alpha}{\partial x_i \partial x_j},
  \label{eq:bkgfd}
\end{equation}
where the subscripts $i,j = 1,2$ denote the spatial indices. Equation \eqref{eq:bkgfd} forms the evolution of the BKG scheme. In the original works~\cite{bardow06,bardow08}, either finite element (FE) or finite difference (FD) is employed to discretize the spatial gradients in Eq.~\eqref{eq:bkgfd}. In Ref.~[14], the central finite-difference scheme on a uniform mesh is used, i.e, the first and second order spatial derivatives are computed as~\cite{rao15},
\begin{subequations}\label{eq:central_schemes}
\begin{align}
   \left. \frac{\partial \tilde f^{+,n}_\alpha}{\partial x_1} \right|_{l,m} &=  \frac{\tilde f^{+,n}_{\alpha,l+1,m} - \tilde f^{+,n}_{\alpha,l-1,m}}{2\Delta x_1},\\
   \left. \frac{\partial \tilde f^{+,n}_\alpha}{\partial x_2} \right|_{l,m} &=  \frac{\tilde f^{+,n}_{\alpha,l,m+1} - \tilde f^{+,n}_{\alpha,l,m-1}}{2\Delta x_2},\\
   \left. \frac{\partial^2 \tilde f^{+,n}_\alpha}{\partial x^2_1} \right|_{l,m}  &=  \frac{\tilde f^{+,n}_{\alpha,l+1,m} + \tilde f^{+,n}_{\alpha,l-1,m} -2 \tilde f^{+,n}_{\alpha,l,m}}{\Delta x_1^2}, \\
   \left. \frac{\partial^2 \tilde f^{+,n}_\alpha}{\partial x^2_2} \right|_{l,m}  &=  \frac{\tilde f^{+,n}_{\alpha,l,m+1} + \tilde f^{+,n}_{\alpha,l,m-1} -2 \tilde f^{+,n}_{\alpha,l,m}}{\Delta x_2^2},\\
   \left. \frac{\partial^2 \tilde f^{+,n}_\alpha}{\partial x_1 \partial x_2} \right|_{l,m} &=
\frac{1}{4\Delta x_1 \Delta x_2}
[\tilde f^{+,n}_ {\alpha,l+1,m+1}
- \tilde f^{+,n}_{\alpha,l-1,m+1}
- \tilde f^{+,n}_{\alpha,l+1,m-1}
+ \tilde f^{+,n}_{\alpha,l-1,m-1}],
   \end{align}
\end{subequations}
where the computational stencil for each node is illustrated in Fig.~\ref{fig:fv_stencil}.
It is noted that if we use the one dimensional Lax-Wendroff scheme to solve Eq.~\eqref{eq:bkgfd} in each discrete velocity direction on a uniform Cartesian grid,
this characteristic based scheme reduces to the Lax-Wendroff LBE scheme developed in~\cite{mcnamara95}.

The Lax-Wendroff scheme Eq.~\eqref{eq:bkgfd} can also be expressed as
\begin{align}
\begin{split}\label{eq:xx1}
\tilde f^{n+1}_\alpha
&= \tilde f^{+,n}_\alpha - \Delta t  \xi_{\alpha i} \frac{\partial}{\partial x_i}\left[\tilde f^{+,n}_\alpha -\frac{\Delta t}{2}\xi_{\alpha j} \frac{\partial \tilde f^{+,n}_\alpha}{ \partial x_j}\right]   \\
&\approx \tilde f^{+,n}_\alpha - \Delta t  \xi_{\alpha i} \frac{\partial}{\partial x_i}\left(\tilde f^{+,n}_\alpha(\bm{x}-\bm{\xi_\alpha}h)\right)\\
&\equiv \tilde f^{+,n}_\alpha - \Delta t  \xi_{\alpha i} \frac{\partial}{\partial x_i}\left(f^{n+1/2}_\alpha(\bm{x})\right).
\end{split}
\end{align}
This means that the BKG scheme can be reformulated as a FV scheme,
\begin{equation}\label{eq:lwfv}
 \tilde f^{n+1}_{\alpha,k} =  \tilde f^{+,n}_{\alpha,k} + \frac{\Delta t}{|V_k|}\mathcal{F}_{\alpha,k,\text{bkg}},
\end{equation}
where
\begin{equation}
  \mathcal{F}_{\alpha,k,\text{bkg}} \equiv  \int_{\partial V_k} (\bm \xi_\alpha \cdot \bm n)  f^{n+1/2}_{\alpha,\text{bkg}} dS,
  \label{eq:G}
\end{equation}
with
\begin{equation}\label{eq:bkg_fb}
  f^{n+1/2}_{\alpha, \text{bkg}}(\bm x_b) = \tilde f^{+,n}_{\alpha}(\bm x_b - h \bm \xi_\alpha)=f^{n}_{\alpha}(\bm x_b - h\bm\xi_\alpha) +  h \Omega\left(f^n_{\alpha}(\bm x_b - h\bm\xi_\alpha)\right).
\end{equation}

To be more specific, we rewrite Eq.~\eqref{eq:bkgfd} in the finite-volume form as
\begin{equation}\label{eq:bkg_fv}
  \tilde f^{n+1}_{\alpha,l,m} = \tilde f^{+,n}_{\alpha,l,m} - \frac{\xi_{\alpha 1} \Delta t}{\Delta x_1} \left[ f^{n+1/2}_{\alpha,l+1/2,m} - f^{n+1/2}_{\alpha,l-1/2,m}\right]
  - \frac{\xi_{\alpha 2} \Delta t}{\Delta x_2} \left[ f^{n+1/2}_{\alpha,l,m+1/2} - f^{n+1/2}_{\alpha,l,m-1/2}\right],
\end{equation}
where $f^{n+1/2}_{\alpha,l\pm 1/2,m}$ and $f^{n+1/2}_{\alpha,l,m \pm 1/2}$ are the DFs at the face-centers of cell $(l,m)$ at the half time step, which depend on interpolation schemes.
If the distribution function is assumed to be a linear piece-wise polynomial, we can obtain the distribution functions at the cell interfaces, e.g.,
\begin{align}
\begin{split}\label{eq:f_left}
  f^{n+1/2}_{\alpha,l-1/2,m} &= \frac{1}{2}[\tilde f^{+,n}_{\alpha,l,m} + \tilde f^{+,n}_{\alpha,l-1,m}] -
  \frac{\xi_{\alpha 1} \Delta t}{2\Delta x_1} [ \tilde f^{+,n}_{\alpha,l,m} - \tilde f^{+,n}_{\alpha,l-1,m} ] \\
&- \frac{\xi_{\alpha 2} \Delta t}{8\Delta x_2} [ \tilde f^{+,n}_{\alpha,l-1,m+1} + \tilde f^{+,n}_{\alpha,l,m+1}
-\tilde f^{+,n}_{\alpha,l-1,m-1} - \tilde f^{+,n}_{\alpha,l,m-1} ]\\
&\approx
\tilde f^{+,n}_{\alpha,l-1/2,m} - \frac{\xi_{\alpha1}\Delta t}{\Delta x_1}\frac{\partial}{\partial x_1}\tilde f^{+,n}_{\alpha,l-1/2,m} - \frac{\xi_{\alpha2}\Delta t}{\Delta x_2}\frac{\partial}{\partial x_2}\tilde f^{+,n}_{\alpha,l-1/2,m}.
\end{split}
\end{align}
One can immediately check the equivalence of Eqs.~\eqref{eq:bkg_fv} and~\eqref{eq:bkgfd}, after calculating the rest flux terms in a similar way like Eq.~\eqref{eq:f_left},
and inserting Eq.~\eqref{eq:central_schemes} to Eq.~\eqref{eq:bkg_fv}.

\subsection{Comparison of the numerical fluxes in the DUGKS and BKG scheme}
\label{sec:numerical}
We now analyse the differences between the DUGKS and the BKG scheme in finite-volume formulation.
This is achieved by analyzing accuracy of the reconstructed distribution function at the cell interface center.
Firstly, it is noted that the exact solution of the DVBE at the cell interface center is
\begin{equation}\label{eq:ana}
  f_{\alpha,\text{exact}}^{n+1/2}(\bm x_b) = f^n_{\alpha}(\bm x_b - h\bm\xi_\alpha) + \int_{0}^{h}\Omega\left(f_\alpha(\bm x_b - h\bm \xi_\alpha + s\bm \xi_\alpha, t^n+s)\right)ds.
\end{equation}
We can immediately find that if we approximate the integration of the collision term in Eq.~\eqref{eq:ana} explicitly,
i.e., assuming $\Omega(\bm x_b - h\bm \xi_\alpha + s\bm \xi_\alpha,s)=\Omega(\bm x_b-h\bm\xi_\alpha,t^n)$,  we get Eq.~\eqref{eq:bkg_fb},
which is the reconstructed distribution function in the BKG scheme.
On the other hand, if we apply trapezoidal rule to the quadrature, we get Eq.~\eqref{eq_charraw}, i.e., the reconstructed cell-interface distribution functions in the DUGKS.
So, in both the DUGKS and the BKG methods, the flux is determined from the local characteristic solution of the DVBE,
and the convection and collision effects are considered simultaneously.
We also indicate that, for those FV/FD schemes that use the simple central difference~\cite{penggw99} or upwind~\cite{guozl03,kim08,patil09} schemes of traditional CFD methods,
 the corresponding cell-interface distribution functions are
\begin{equation}
  f^{n+1/2}_{\alpha,\text{central}}(\bm x_b) = f^{n}_{\alpha}(\bm x_b)
  \label{eq:cfd1}
\end{equation}
and
\begin{equation}
  f^{n+1/2}_{\alpha,\text{upwind}}(\bm x_b) = f^{n}_{\alpha}(\bm x_b - h\bm\xi_\alpha),
  \label{eq:cfd2}
\end{equation}
where the collision effect is totally ignored.

The effects of the different treatments of the integration of the collision term in Eq.~\eqref{eq:ana}
can be analyzed by using the Chapman-Enskog expansion method~\cite{ohwada02}.
By approximating the distribution function by its first-order Chapman-Enskog solution,
\begin{equation}\label{eq:ce}
  f_{\alpha }=f_{\alpha}^{(0)}+\tau f_{\alpha }^{(1)}+O({{\tau }^{2}}),
\end{equation}
with $f_{\alpha }^{(0)}=f_{\alpha }^{eq}$, we have
\begin{multline}\label{eq:ana_ce}
     f_{\alpha,\text{exact}}^{n+1/2}(\bm x_b)  = f^{(0),n}_{\alpha}(\bm x_b - h\bm \xi_\alpha ) + \tau f^{(1),n}(\bm x_b - h\bm \xi_\alpha)
 - \int_0^h {f^{(1)}_\alpha}({\bm x_b} - h\bm\xi_\alpha  + \bm \xi_\alpha s,s) ds + O(\tau^2).
\end{multline}
Therefore, for continuous flows where $\tau\ll h$,
the distribution function at the cell interface reconstructed in either the BKG or the DUGKS method is a second-order approximation of the exact one.
Furthermore, with the trapezoidal rule for the collision term, the DUGKS is expected to be more accurate than the BKG method which uses a lower order explicit rule.
In regions with large velocity gradients, where the collision effect or $f^{(1)}$ is important, the BKG scheme may yield significant error.
More importantly, as the DUGKS employs an implicit treatment of the collision term in the evaluation of numerical flux, it is expected to be more stable than the BKG scheme.

Finally, we shall remark that if the distribution functions at cell interfaces are obtained by direct interpolations,
i.e., neglecting the integral of the collision term in Eq.~\eqref{eq:ana}, the leading error of the approximation is $-hf_\alpha^{(1),n}$.
As the integral of $f^{(1),n}$ on right hand side of Eq.~\eqref{eq:ana_ce} contributes to the diffusive flux,
the lack of the collision term is equivalent to introduce a numerical viscosity proportional to $\Delta t$~\cite{ohwada02,chensz15,xuk01,xuk01b}.
This explains why many other FV based lattice Boltzmann schemes have to keep the time step much smaller than the collision time to obtain accurate results.

\subsection{No-slip boundary condition}
\label{sec:wallboundary}
\begin{figure}[htbp]
\centering
\includegraphics[width=0.5\textwidth]{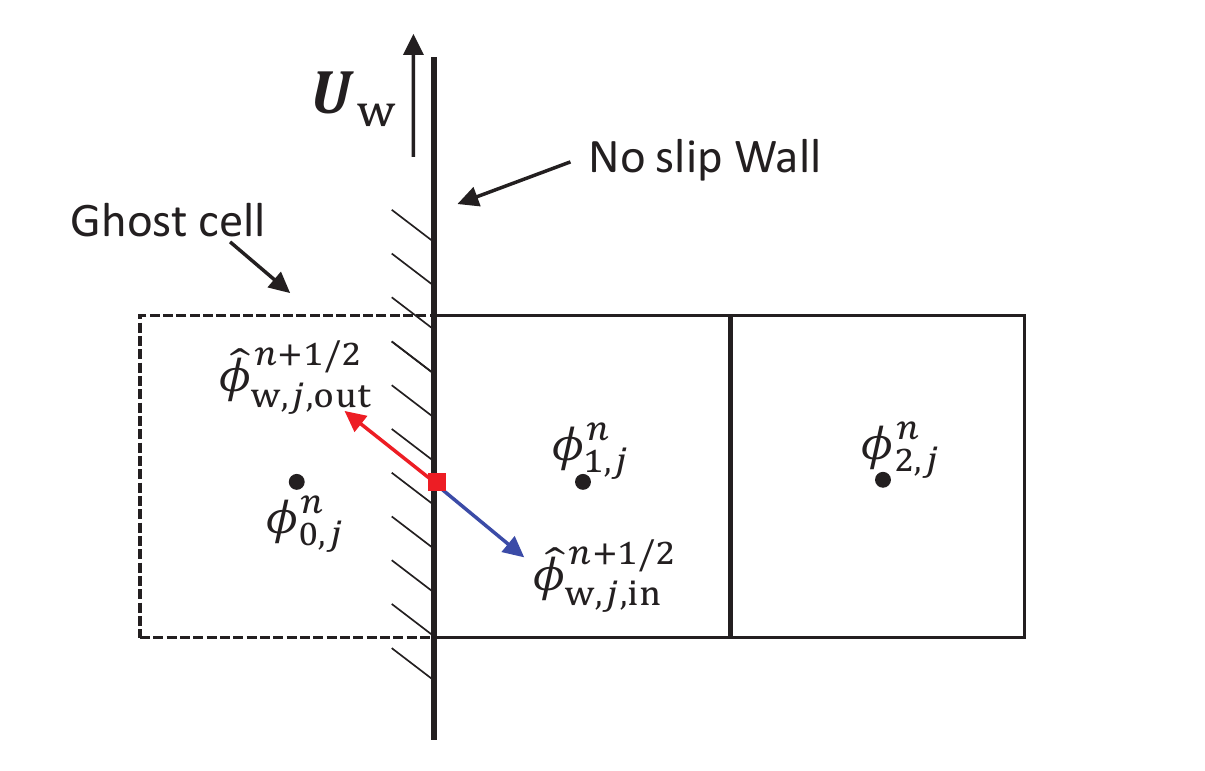}
\caption{
Implementation of no-slip boundary condition
} \label{fig:boundary_condition}
\end{figure}

In this subsection, we briefly mention the implementation of no-slip boundary condition for the BKG scheme and DUGKS methods,
The basic idea here is to mimic the half-way bounce-back rule of the standard LBM~\cite{guobook} by reversing the DFs at boundary faces at middle time steps.
Fig.~\ref{fig:boundary_condition} illustrates a boundary face located at a no slip wall with velocity $\bm U_\text{w}$.
Both of the incoming and outgoing DFs at the boundary face at middle time steps have to be provided to update the cell-centered DFs.
We denote the incoming and outgoing DFs by $\hat{\phi}_\text{w,in}^{n+1/2}$ and $\hat{\phi}_\text{w,out}^{n+1/2}$ respectively,
where $\hat{\phi}_\text{w}^{n+1/2}$ stands for $\bar{f}_\text{w}^{n+1/2}$ in the DUGKS and $f_\text{w}^{n+1/2}$ in BKG scheme.
The ghost cell method is used to facilitate the implementation of the no-slip boundary condition.
An extra layer of cells (ghost cells) are allocated outside of the wall.
The unknown DFs $\phi^n$ in the ghost cells are extrapolated linearly from the cell centers of the neighboring inner cells.
Here, $\phi^n$ stands for $\bar{f}^{+,n}$ in the DUGKS and $\tilde{f}^{+,n}$ in the BKG scheme.
Then we can compute the $\hat{\phi}_\text{w,out}^{n+1/2}$ normally.
After that, the incoming DFs are calculated in the same way as the half-way bounce-back rule in the standard LBM~\cite{guobook},
\begin{equation}
    \hat{\phi}_{\text{w},\bar\alpha}^{n+1/2}= \hat{f}_{\text{w}, \alpha}^{n+1/2} - 2w_\alpha \frac{\bm \xi_\alpha \cdot \bm U_\text{w}}{c_s^2},
\end{equation}
where $\alpha$ stands for an outgoing DF direction and $\bar \alpha$ is its reverse direction.


\section{Numerical tests}
In this section, we compare the DUGKS and the BKG scheme in terms of accuracy, stability and computational efficiency by simulating several two dimensional flows.
The first one is the unsteady Taylor-Green vortex flow which is free from boundary effect, and exhibits an analytical solution exists for this problem,
the second test case is the lid-driven cavity flow, which is used to evaluate the accuracy and stability,
and the last one is the laminar boundary layer flow problem, which is used to verify the dissipation property of the the DUGKS and the BKG methods.
In all of our simulation, $c_s$ is set to be $1/\sqrt{3}$ and the CFL number is set to be 0.5 unless stated otherwise.

\subsection{Taylor-Green vortex flow}
This problem is a two dimensional unsteady incompressible flow in a square domain with periodical condition in both directions.
The analytical solution is given by
\begin{subequations}
  \begin{align}
u(x,y,t) =  & - U_0\cos(2\pi x)\sin(2\pi y)\exp(-8\pi^2 \nu t),\\
v(x,y,t) =  &  U_0\sin(2\pi x)\cos(2\pi y) \exp(-8\pi^2 \nu t),\\
p(x,y,t) =  &   -\frac{U_0^2}{4} \left[ \cos(4\pi x) + \cos(4\pi y) \right] \exp(-16\pi^2\nu t),
\end{align}
\label{eq:tvana}
\end{subequations}
where $U_0$ is a constant indicating the kinetic energy of the initial flow field, $\nu$ is the shear viscosity, $\bm u = (u,v)$ is the velocity, and $p$ is the pressure.
The computation domain is $0<x<L$ and $0<y<L$ with $L=1$.
We set $U_0=1/\sqrt{3}\times 10^{-2}$ and $\nu=1/\sqrt{3}\times 10^{-4}$.
The corresponding Reynolds number and Mach number are $\text{Re}=U_0 L/\nu = 100$ and $\text{Ma}=U_0/c_s=0.01$, respectively.
The initial distribution function is computed from the Chapman-Enskog expansion at the Navier-Stokes order~\cite{guobook}
\begin{equation}
  f_\alpha(\bm x, 0) = f^{\text{eq}}_\alpha - \tau [\partial_t f^{\text{eq}}_\alpha
+  \bm \xi_\alpha \cdot \bm \nabla_x f^{\text{eq}}_\alpha],
  \label{}
\end{equation}
where the equilibrium distribution functions are evaluated from the initial analytical solution.

We first evaluate the spatial accuracy of the DUGKS and BKG scheme by simulating the flow with varies mesh resolutions ($N\times N$).
As we are analyzing the spatial accuracy, the time step is set to a very small value ($\Delta t=2\tau$) to suppress the errors caused by the time step size.
The $L_2$-error of the velocity filed is measured,
\begin{equation}
  E_u(t) = \frac{\sqrt{ \sum_{x,y}|\bm u_n(x,y,t) - \bm u_a(x,y,t)|^2}}{\sqrt{ \sum_{x,y}|\bm u_a(x,y,t)|^2}},
  \label{}
\end{equation}
where $\bm u_a$ and $\bm u_n$ are the analytical solution and numerical solution respectively.
The $L_2$-errors at the half-life time $t_c=\ln(2)/(8\nu\pi^2)$  of the two schemes are measured and listed in Table~\ref{tab:tv_Nerr}.
The results are listed in Table~\ref{tab:tv_Nerr}.
It can be seen that both of the methods are of second order accuracy in space.
But the errors computed from DUGKS results are smaller than those of the BKG scheme on the same mesh resolutions.
\begin{table}[htbp]
  \centering
  \caption{$L_2$-errors of the velocity filed for the Taylor-Green vortex flow}
    \begin{tabular}{lrrrrr}
    \toprule
& N     & 16    & 32    & 64    & 128 \\
    \midrule
    \multicolumn{1}{l}{\multirow{2}[0]{*}{DUGKS}} & $E_u(t_c)$ & 4.1416E-03 & 1.0852E-03 & 2.6829E-04 & 6.1103E-05 \\
    \multicolumn{1}{l}{} & order & -     & 1.93  & 2.02  & 2.13  \\
    \midrule
    \multicolumn{1}{l}{\multirow{2}[0]{*}{BKG scheme}} & $E_u(t_c)$ & 1.7025E-02 & 4.3950E-03 & 1.1015E-03 & 2.6945E-04 \\
    \multicolumn{1}{l}{} & order & -     & 1.95  & 2.00  & 2.03  \\
    \bottomrule
    \end{tabular}%
  \label{tab:tv_Nerr}%
\end{table}%

\begin{figure}[htbp]
\centering
\includegraphics[width=0.5\textwidth]{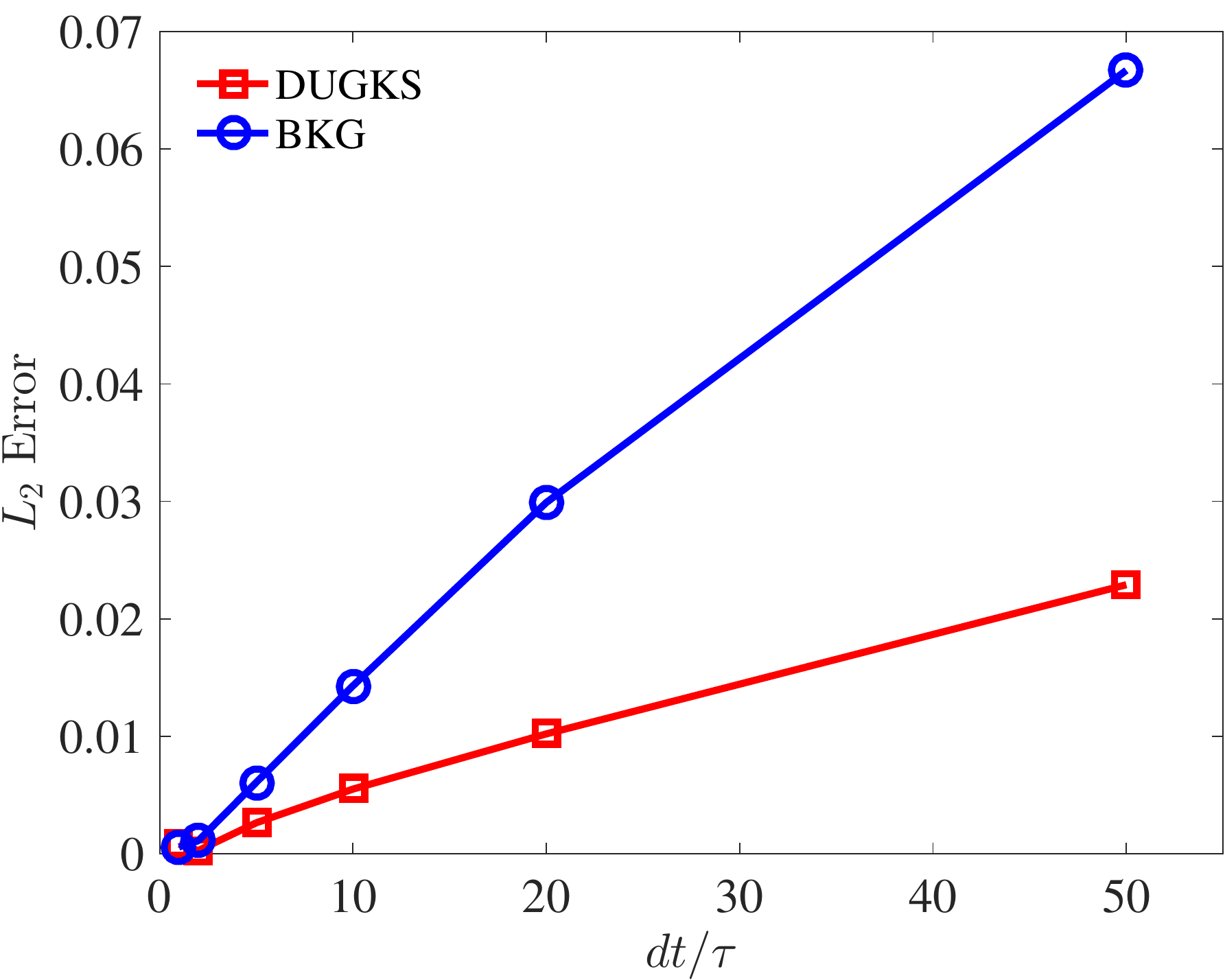}
\caption{
    Errors of the velocity field at $t_c$ using varies $dt/\tau$ on a $64\times 64$ mesh.
} \label{fig:tv_dtErr}
\end{figure}

\begin{figure}[htbp]
\centering
\includegraphics[width=0.5\textwidth]{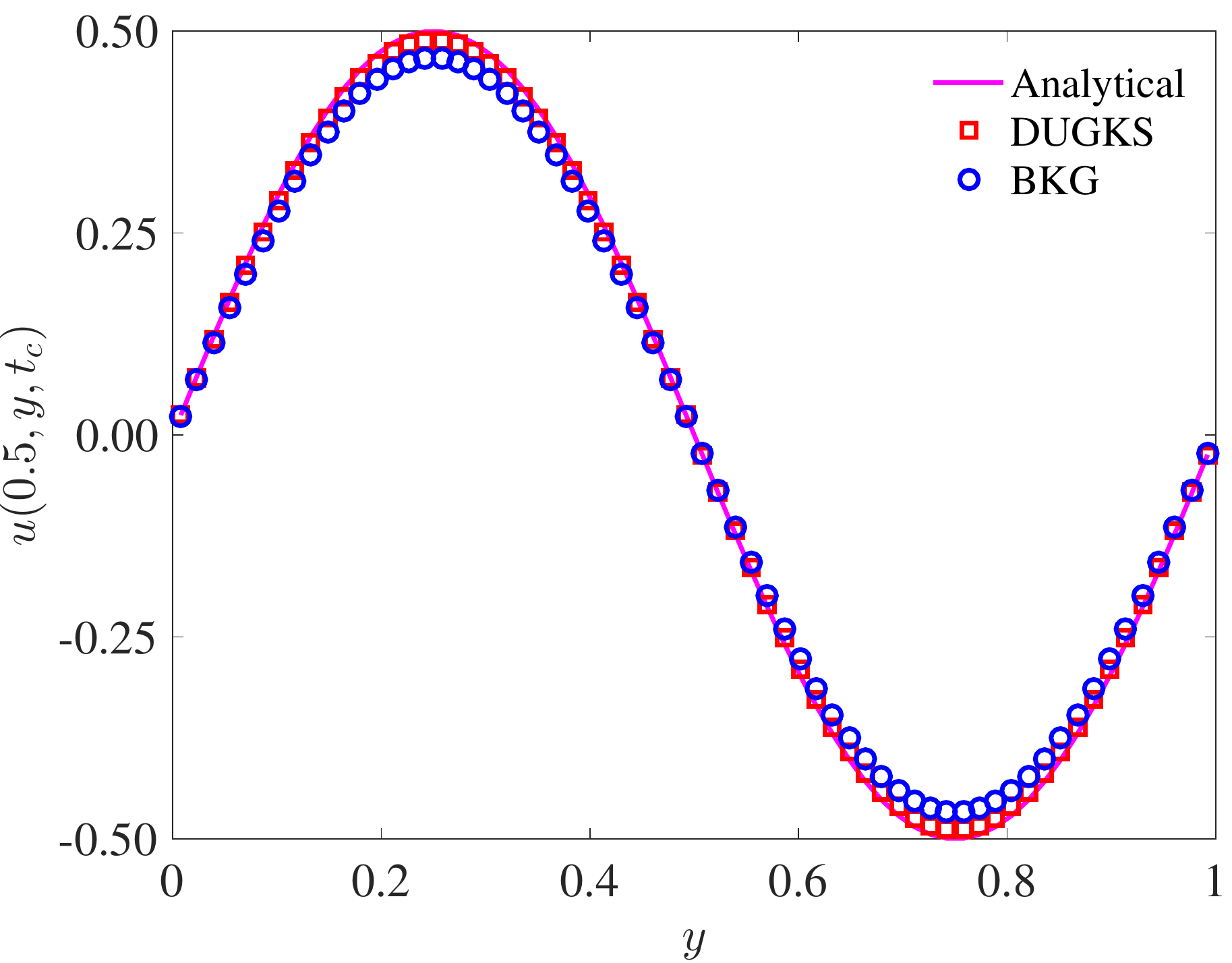}
\caption{
Velocity profile along the line $x=0.5$ at $t_c$ on a $64\times 64$ mesh, the time step is $\Delta t= 50\tau $.
} \label{fig:tv_uprofile}
\end{figure}

%
%
Since Both the DUGKS and BKG methods can admit a time step larger than the relaxation time,
we now investigate their performance at large values of $\Delta t/\tau$ .
We fix the mesh size ($N=64$) and the relaxation time but change the time step.
The $L_2$-errors at $t_c$ are shown in Fig.~\ref{fig:tv_dtErr}, from which
we can see that the errors scale almost linearly with the time step size for both methods.
Particularly, the two methods still give reasonably accurate results $\Delta t/\tau$ is as large as 50, as shown in Fig.~\ref{fig:tv_uprofile}.
And again, the errors of the DUGKS are smaller than those of the BKG scheme in all cases.
The two methods both blow up as $\Delta t/\tau=100$ since the CFL number goes beyond 1 at this condition.

Here, we also discuss the computational efficiencies of the DUGKS and the BKG scheme
when implementing both of the schemes in FV framework, their only difference is the computing of numerical flux.
The DUGKS introduces two sets of additional DFs and macro variables at cell faces are required.
So it can be expected that the DUGKS's computing cost is obviously higher than that of the BKG scheme.
For example on an Intel Xeon E5-2670v3@2.6GHz CPU,
the computation times for 10,000 steps of the BKG and the DUGKS with $64\times 64$ mesh are 10.5s and 19.7s respectively,
meaning the DUGKS is about one time more expansive than the BKG scheme.

\subsection{Lid-driven cavity flow}
Incompressible two dimensional lid-driven cavity flow is a popular benchmark problem for numerical schemes.
Here, we use it to evaluate the accuracy and stability of the two schemes at different Reynolds numbers.
The flow domain is a square cavity with length $L$.
The top wall moves with a constant velocity $U_w$, while other walls are kept fixed.
The Reynolds number is defined as $\text{Re} = U_w L /\nu$ with $\nu$ being the viscosity of the fluid.
In the computation, we set $L=1$, $U_w = 0.1$, and the viscosity of the fluid is adjusted to achieve different Reynolds numbers.
Uniform Cartesian meshes with grid number $N\times N$ are used in our simulations.

\begin{figure}[p]
\centering
\subfloat[BKG scheme]{\includegraphics[width=0.44\textwidth]{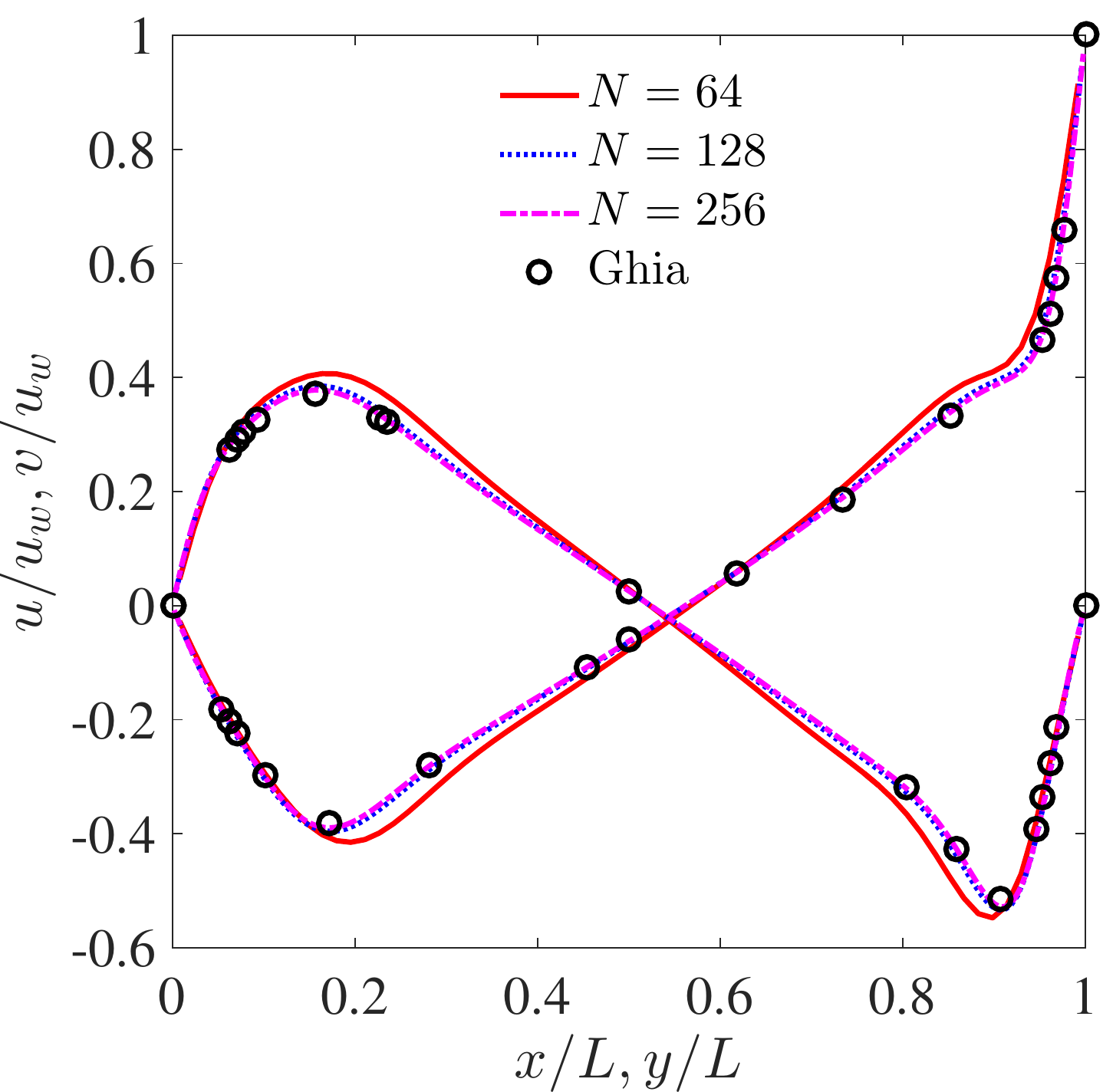}}
\quad
\subfloat[DUGKS]{\includegraphics[width=0.44\textwidth]{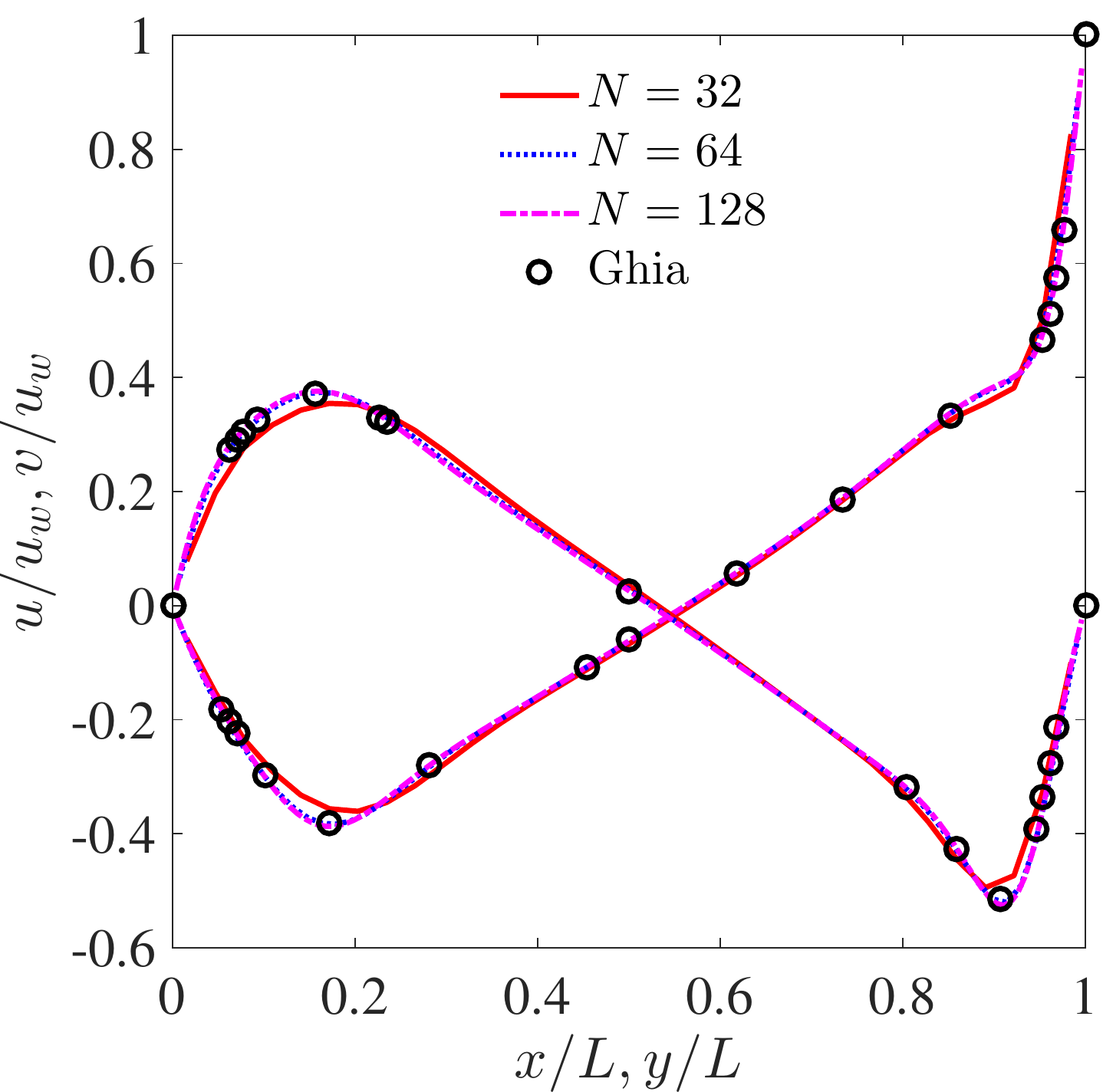}}
\caption{
  Velocity profiles along the cavity center lines at $\text{Re}=1000$ with different mesh resolutions.
} \label{fig:cavity_Re1000}
\end{figure}

\begin{figure}[p]
\centering
\subfloat[BKG scheme]{\includegraphics[width=0.44\textwidth]{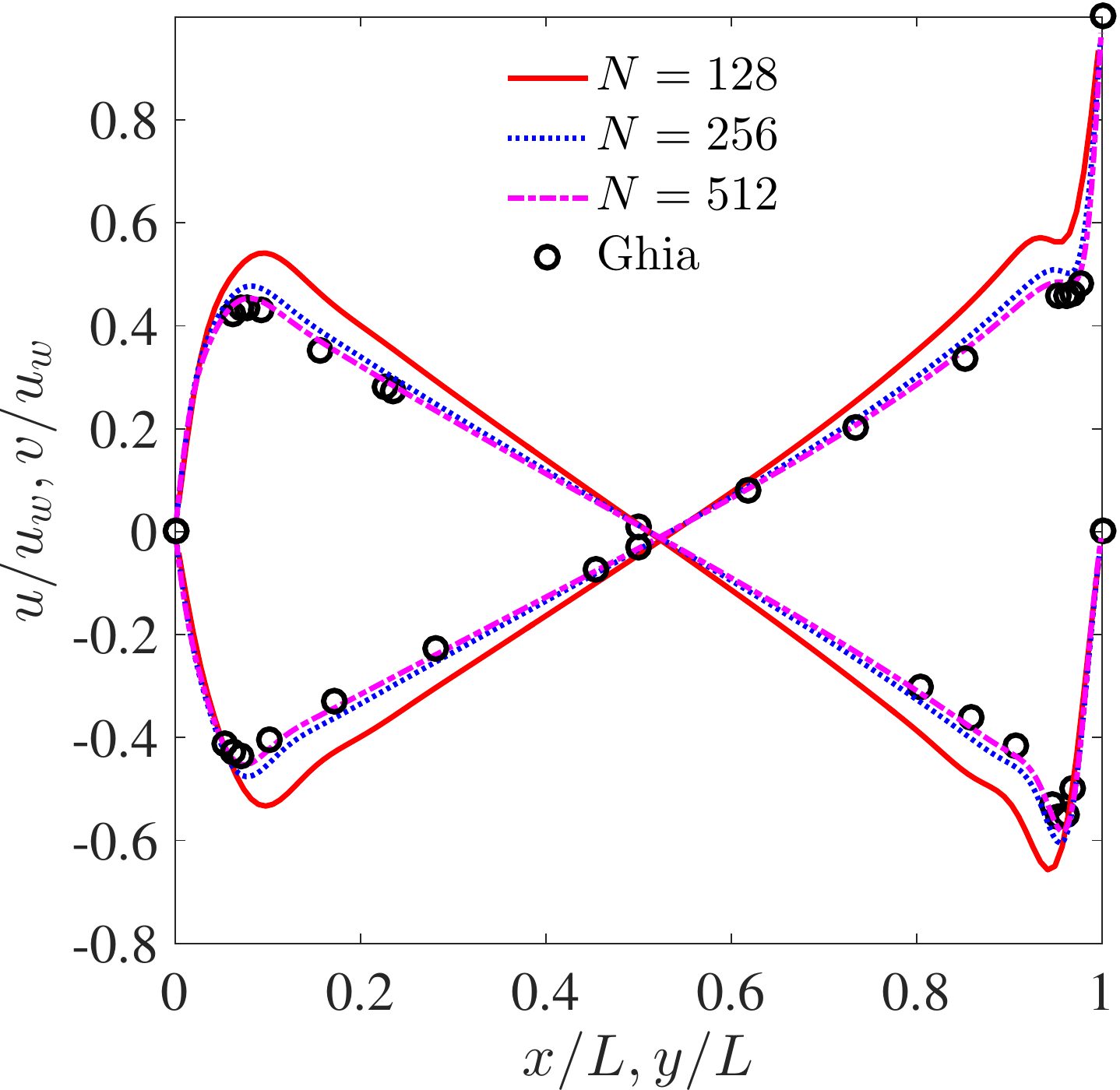}}
\quad
\subfloat[DUGKS]{\includegraphics[width=0.44\textwidth]{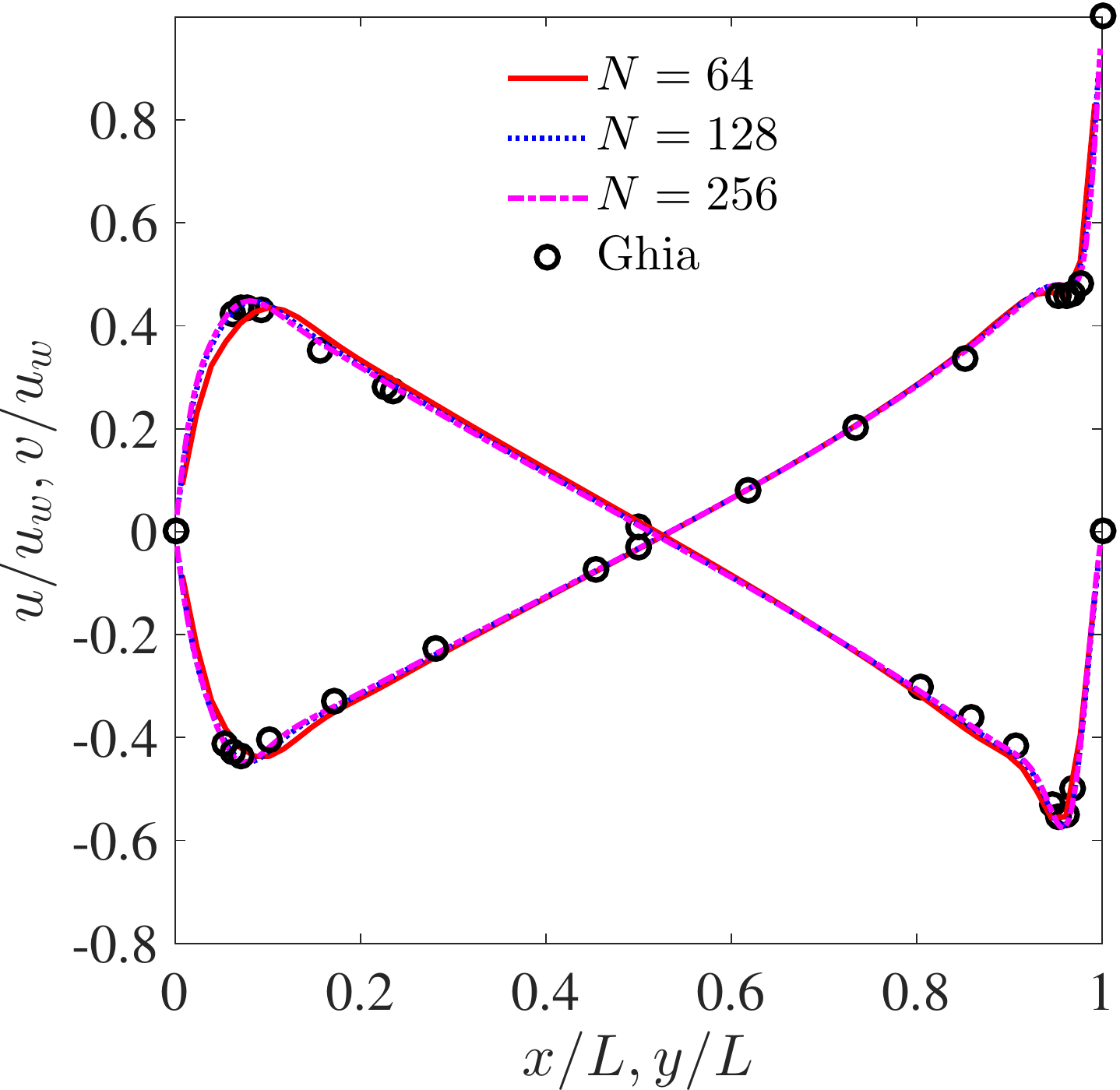}}
\caption{
  Velocity profiles along the cavity center lines at $\text{Re}=5000$ with different mesh resolutions.
} \label{fig:cavity_Re5000}
\end{figure}

\begin{figure}[p]
\centering
\subfloat[BKG scheme]{\includegraphics[width=0.44\textwidth]{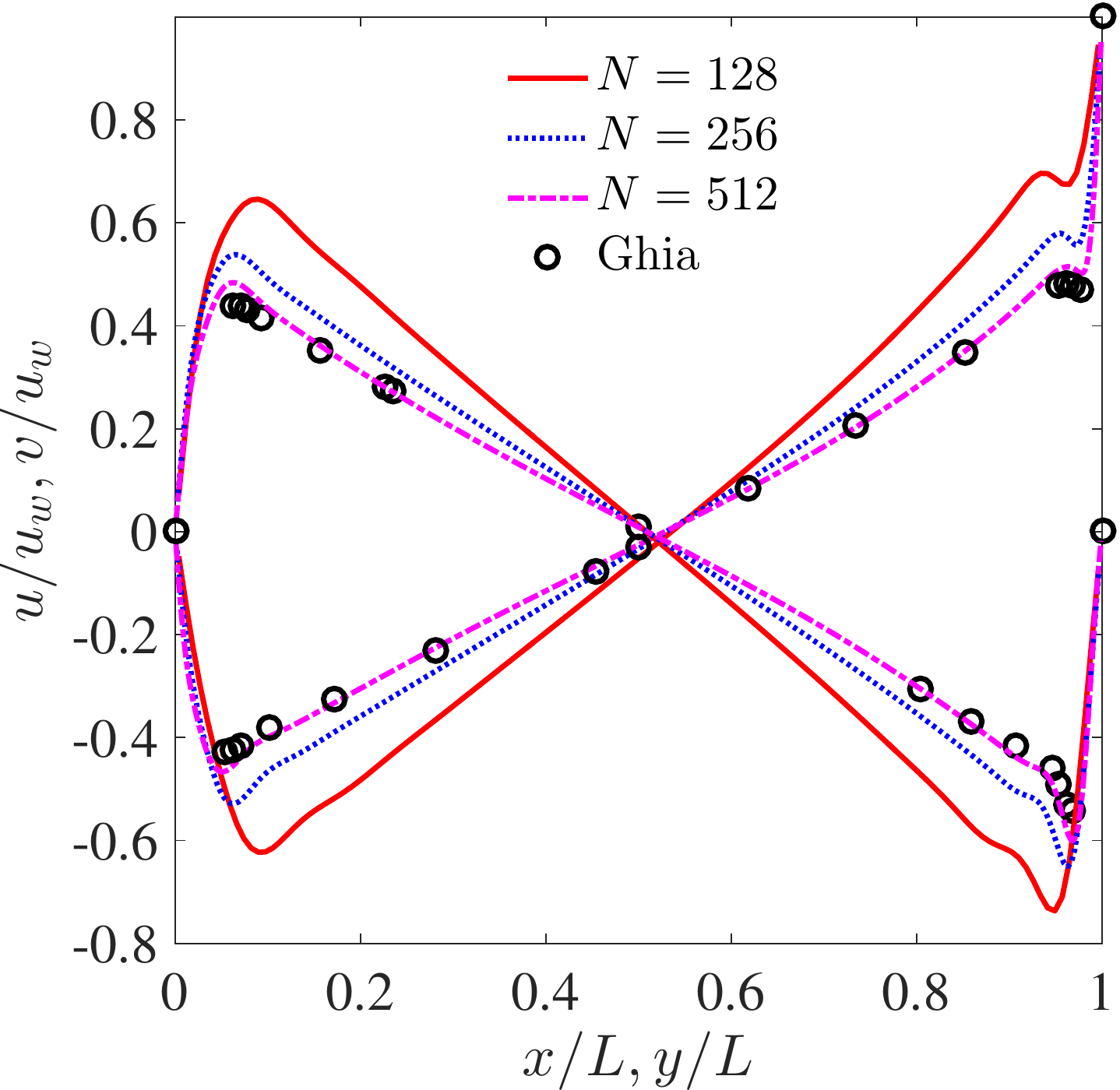}}
\quad
\subfloat[DUGKS]{\includegraphics[width=0.44\textwidth]{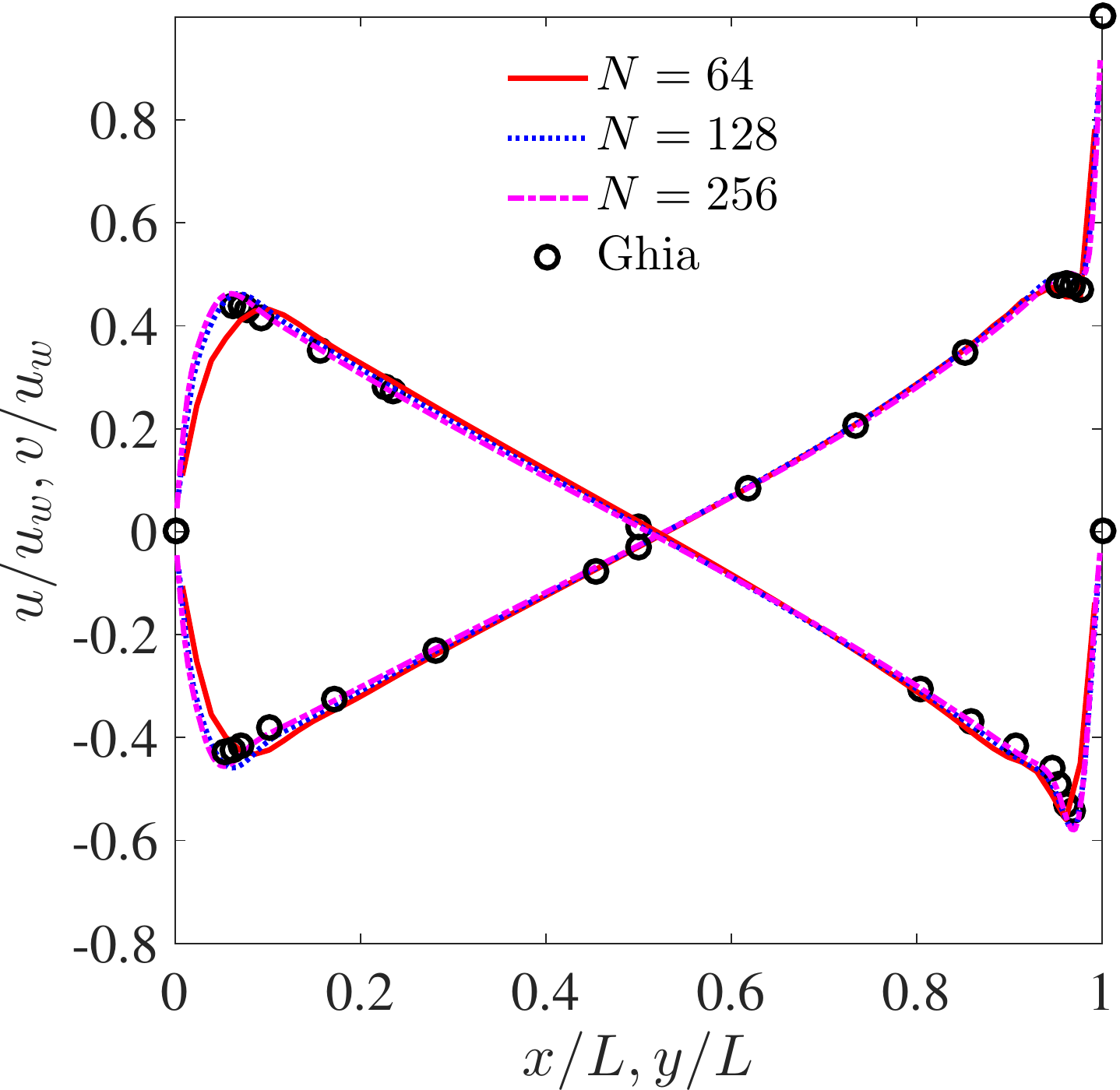}}
\caption{
  Velocity profiles along the cavity center lines at $\text{Re}=10000$ with different mesh resolutions.
} \label{fig:cavity_Re10000}
\end{figure}

We first simulate the flow at $\text{Re}=1000, 5000$ and $10000$ with different mesh resolutions to compare the accuracy and stability of the DUGKS and BKG methods.
The velocity profiles at steady states along the vertical and horizontal center lines
predicted by the two schemes are presented in Figs.~\ref{fig:cavity_Re1000}-\ref{fig:cavity_Re10000}.
The benchmark solutions~\cite{ghia82} are also included for comparison.
It should be noted that, the grid numbers used in the BKG schemes are doubled from those in the DUGKS at each Reynolds number, in that
the BKG computations are unstable at the coarsest meshes used in the DUGKS.
From these results, we can clearly observe that the DUGKS scheme gives more accurate results than the BKG scheme, especially at large Reynolds numbers.
Furthermore, the results show that the DUGKS is insensitive to mesh resolutions, while the BKG scheme is rather sensitive.
Generally, much finer meshes should be used in the BKG scheme to obtained accurate results.
Specifically, the horizontal velocity profiles in the boundary layer of the top wall departure from the benchmark solutions severely at high Reynolds numbers (see Figs.~\ref{fig:cavity_Re5000} and~\ref{fig:cavity_Re10000}) with coarser meshes,
which was also observed in~\cite{rao15}.
Contrary to the BKG scheme, the DUGKS gives surprisingly good results with the same meshes even at $\text{Re} = 10000$.

\begin{figure}[htbp]
\centering
\subfloat[]{\includegraphics[width=0.44\textwidth]{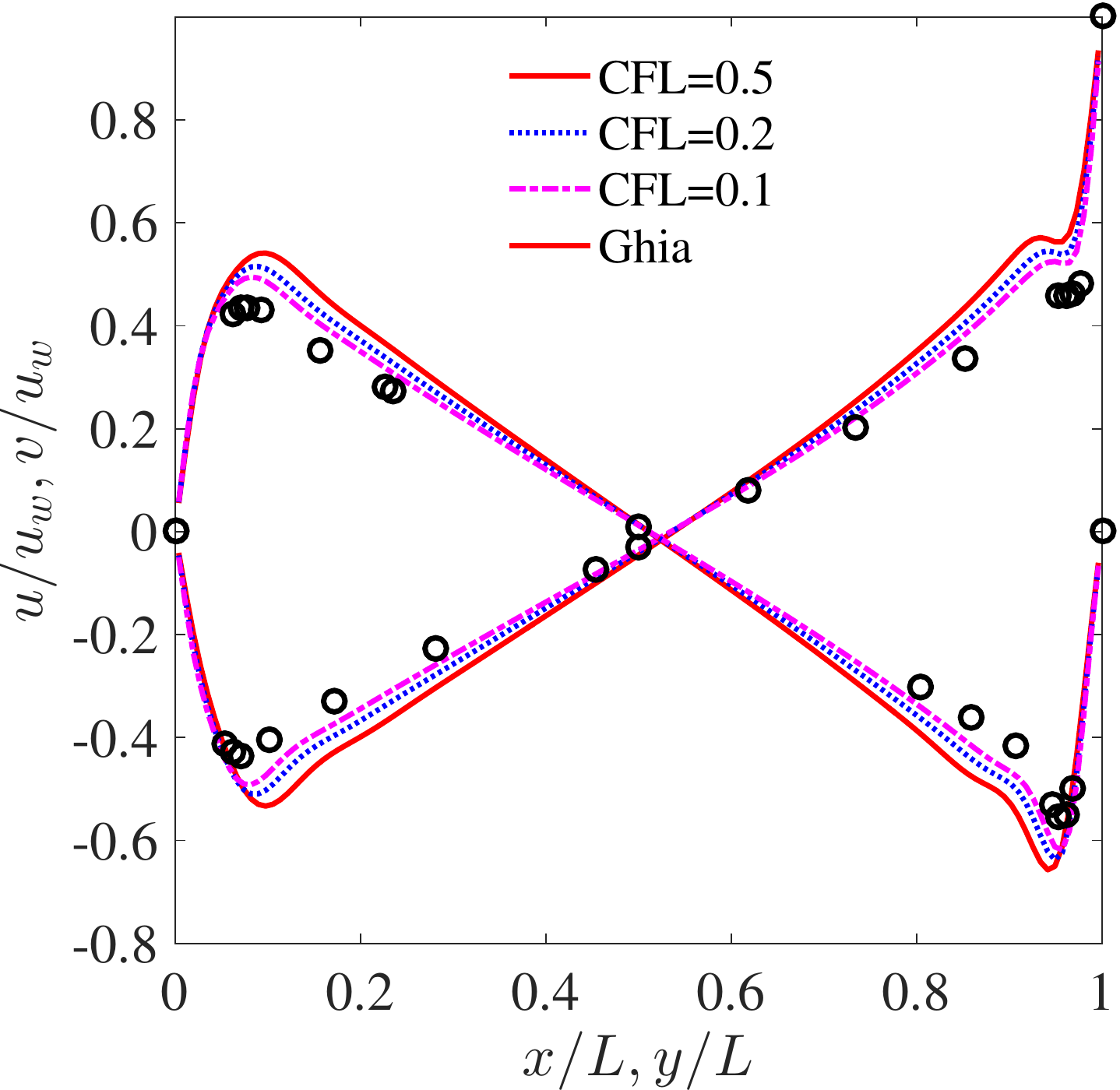}}
\quad
\subfloat[]{\includegraphics[width=0.44\textwidth]{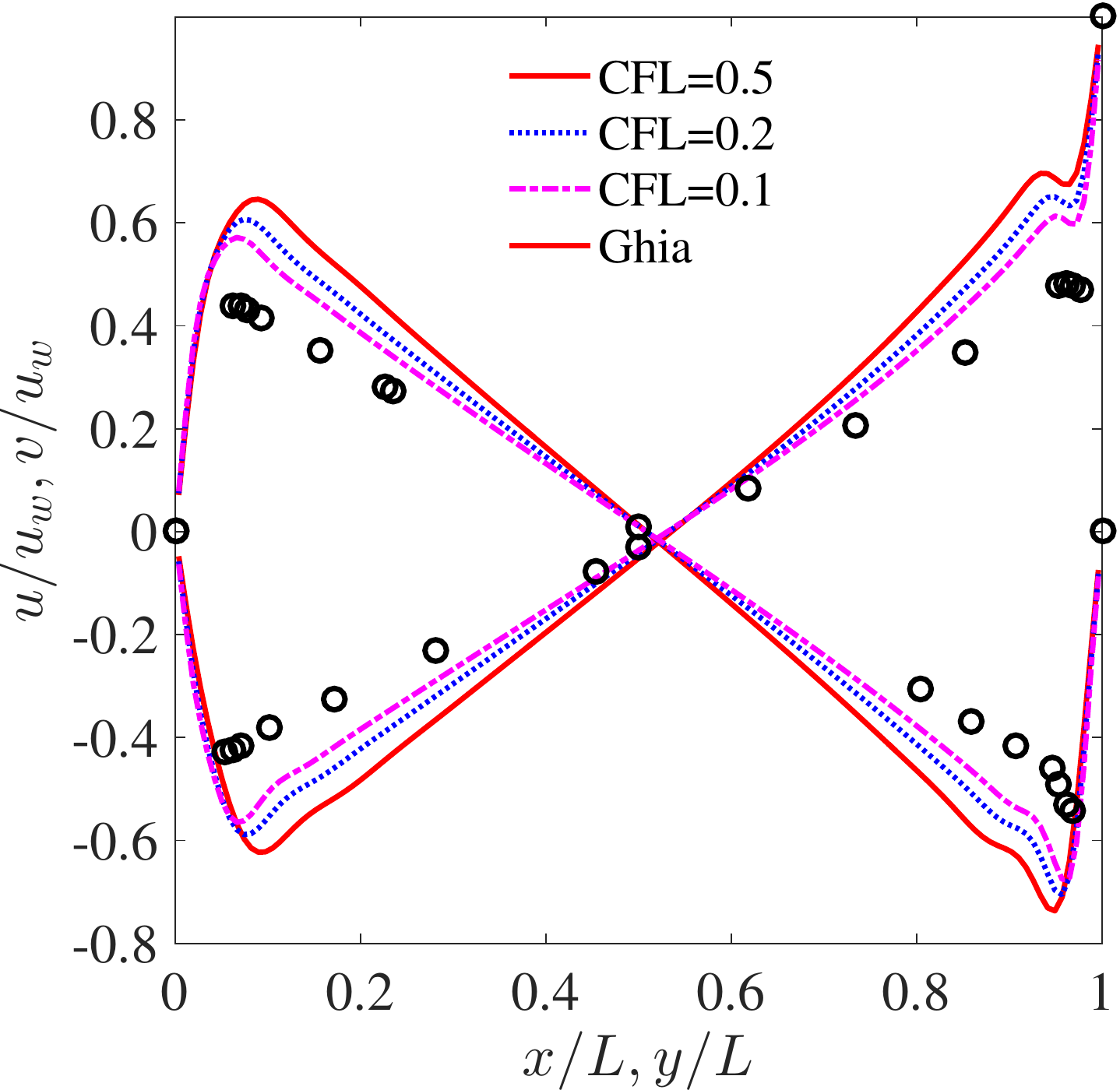}}
\caption{
  Velocity profiles predicted by the BKG scheme at with different CFL numbers on a $128\times 128$ mesh. (a) $\text{Re}=5000$, (b) $\text{Re}=10000$.
} \label{fig:cavity_cfl}
\end{figure}

As has been analyzed in Sec.~\ref{sec:numerical},
the only difference between the DUGKS and the BKG scheme is the treatment of the quadrature for the collision term
in the reconstruction of the cell-interface distribution function, and the difference scales with the time step,
which have been confirmed in the test of Taylor-Green vortex flow.
Now we explore the effect of time step on the solution of this steady flow for the BKG scheme.
We simulate the flow at $\text{Re}=5000$ and $10000$ using various CFL numbers with a fixed grid ($N=128$).
The calculated velocity profiles are shown in Fig.~\ref{fig:cavity_cfl}.
We can see that the errors decrease with decreasing CFL number.
But even with CFL=0.1, the errors are still much larger than those of the DUGKS.

\begin{figure}[htbp]
  \centering
  \includegraphics[width=0.6\textwidth]{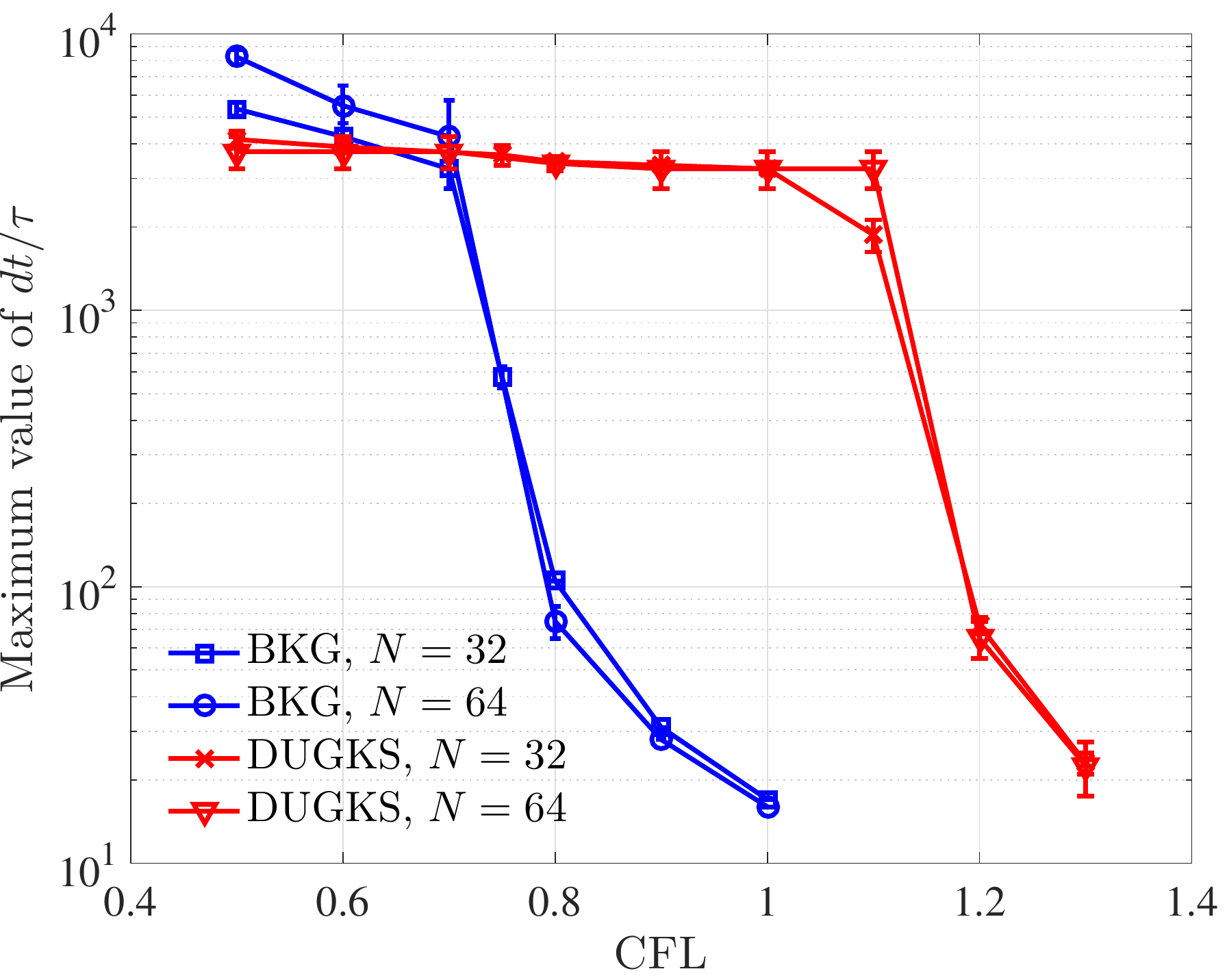}
  \caption{Maximum values of  $\Delta t/\tau$ for stable computations of the cavity flow.}
  \label{fig:cavity_stab}
\end{figure}

We also use the cavity flow to assess and compare the stability of the two schemes.
Generally, the stability of the numerical schemes for the BGK equation is affected by the treatments of both the advection term and the collision term.
The stability for an explicit discretization of the advection term is controlled by the CFL number,
while the stability due to the collision term treatment depends on the ratio of $\Delta t$ and the collision time $\tau$, i.e., $\Delta t/\tau$.
The maximum values of $\Delta t/\tau$ at varies CFL numbers for a stable computation on the $32\times 32$ and $64\times 64$ meshes are measured
and presented in Fig.~\ref{fig:cavity_stab} with error ranges.
It can be seen that there exists a clear distinction between the BKG and the DUGKS methods.
For the BKG scheme, the computation is unstable at moderately large $\Delta t/\tau$ even though $\text{CFL}<1$,
while for the DUGKS, the stability is almost not affected by the CFL number as long as $\text{CFL}<1.1$.
This observation confirms to the analysis in Sec.~\ref{sec:numerical} that the numerical stability is also affected by the treatment of the collision term in the evaluation of numerical flux.
Computing the collision term implicitly both in Eq.~\eqref{eq_raw} and Eq.~\eqref{eq_charraw} makes the DUGKS a rather robust scheme.

\subsection{Laminar boundary layer over a flat plate}
\begin{figure}[htbp]
  \centering
  \includegraphics[width=0.8\textwidth]{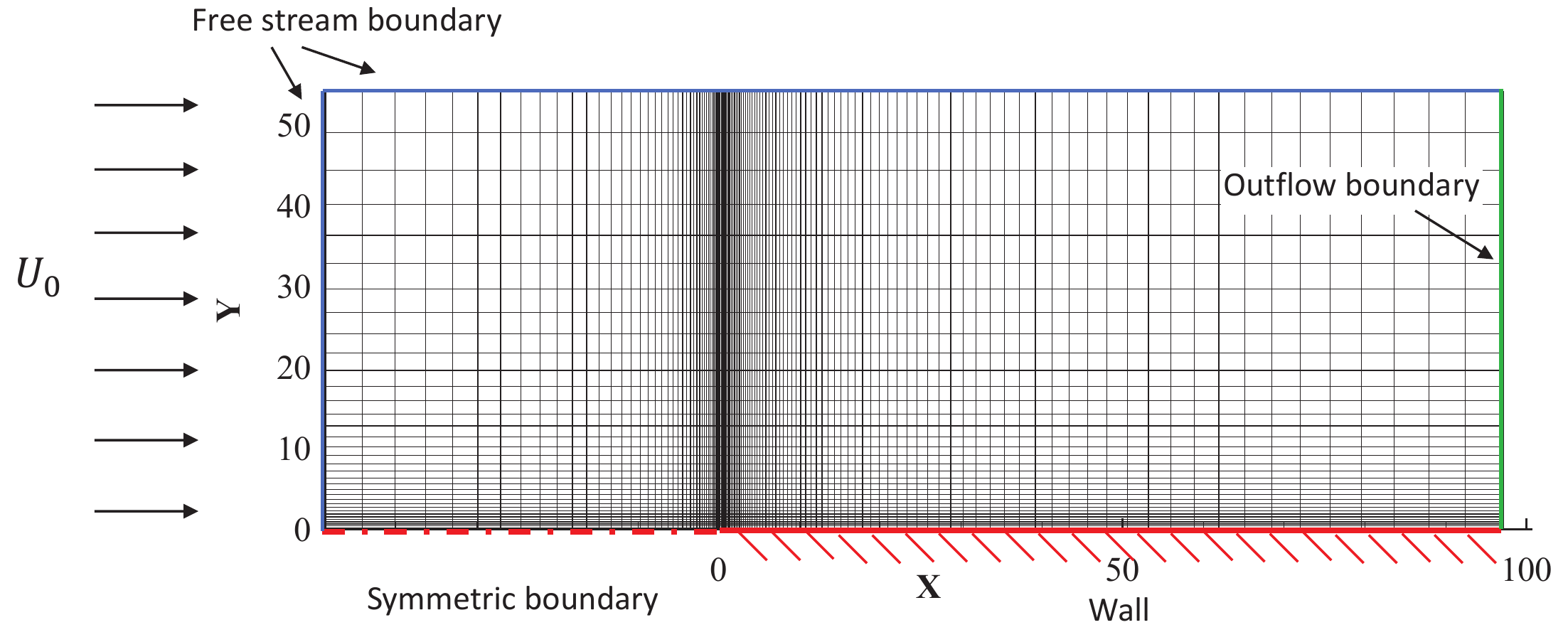}
  \caption{Mesh for the laminar boundary layer.}
  \label{fig:bl_mesh}
\end{figure}

In the cavity flow, it is observed that the BKG scheme fails to capture the boundary layer accurately near the top wall of the cavity for large Reynolds numbers.
In this subsection, we use the laminar flow over a flat plate as a stand-alone case to check this phenomenon and therefore,
evaluate the dissipation characteristics of the BKG scheme and the DUGKS.
The flow configuration of this problem is sketched in Fig.~\ref{fig:bl_mesh}.
A uniform flow with horizontal velocity $U_0$ flows past a flat plate with length $L$.
This steady problem has an analytical self-similar Blasius solution.
The Reynolds number is defined as $\text{Re} = U_0 L /\nu$, where $\nu$ is the kinematic viscosity.
In the simulations, we set $U_0=0.1$, $L=94.76$ and $\text{Re}=10^5$.
The boundary layer is very thin at such a high Reynolds number,
so non-uniform structured meshes stretched in the vertical direction are employed~(Fig.\ref{fig:bl_mesh}).
The cell size along $Y$ direction increases with a ratio $A_y=1.1$,
and the height of the first layer is $\Delta y_{\text{min}}$.
The grid number in the $Y$ direction is adjusted according to $\Delta y_{\text{min}}$ to make sure the height of the computation domain is right beyond 50.
The cell size in the $X$ direction is refined at the leading edge of the plate, with $\Delta x_{\text{min}} = 0.1$ to account for the singularity of the flow behavior there.
The increasing ratios of the cell size to the downstream and upstream from the leading edge are $A_r=1.05$ and $A_l=1.1$, respectively.
The total cell number in the $X$ direction is 120, with 80 cells distributed on the plate.
Free stream condition is applied to the left and top boundaries.
Outflow boundary condition is applied to the right boundary, and symmetric boundary condition is used at the section before the plate at the bottom boundary.
No-slip boundary condition is imposed at the bottom wall and is realized by the method described in Sec.~\ref{sec:wallboundary}.

\begin{figure}[htdp]
\centering
\subfloat[]{\includegraphics[width=0.48\textwidth]{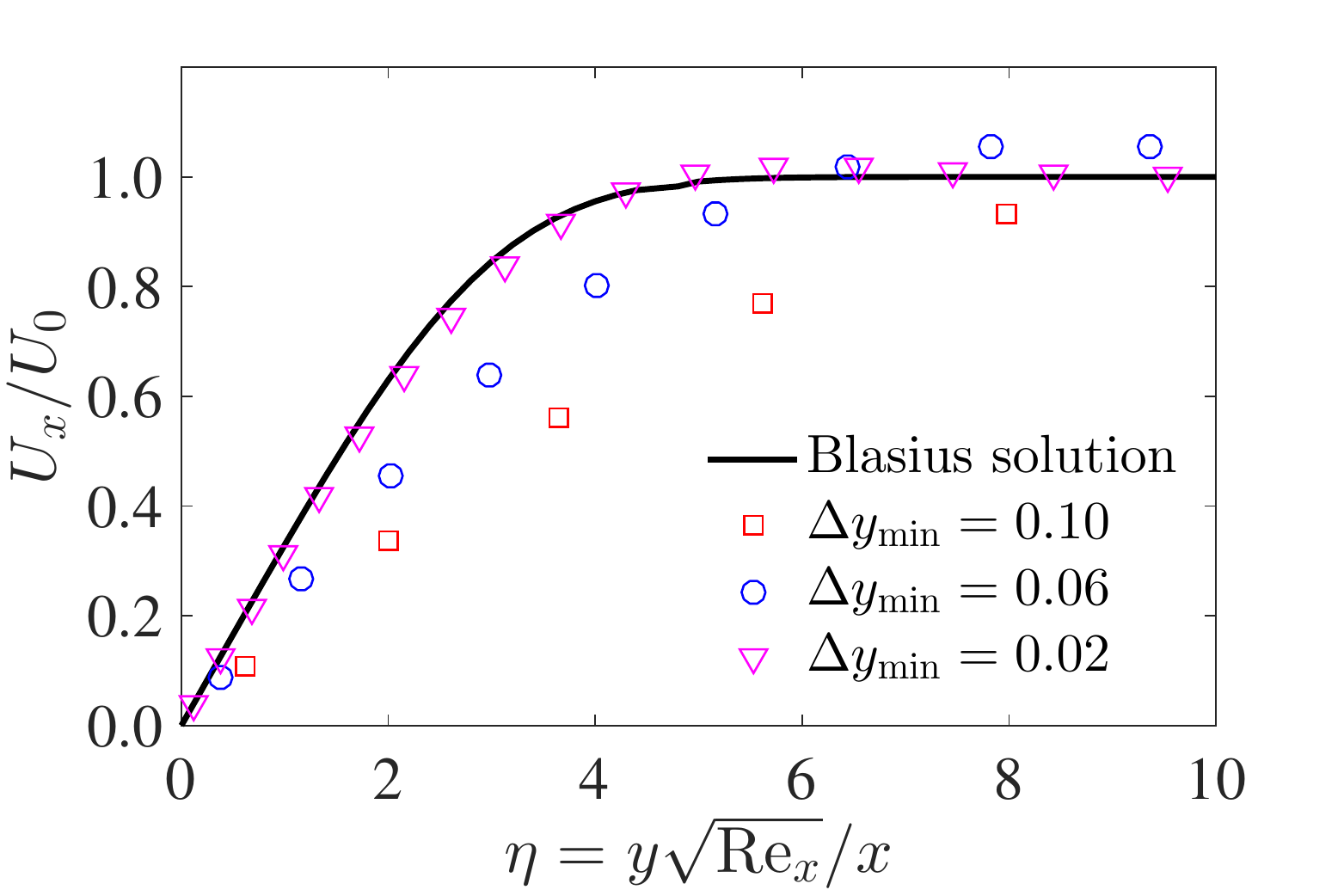}}
\quad
\subfloat[]{\includegraphics[width=0.48\textwidth]{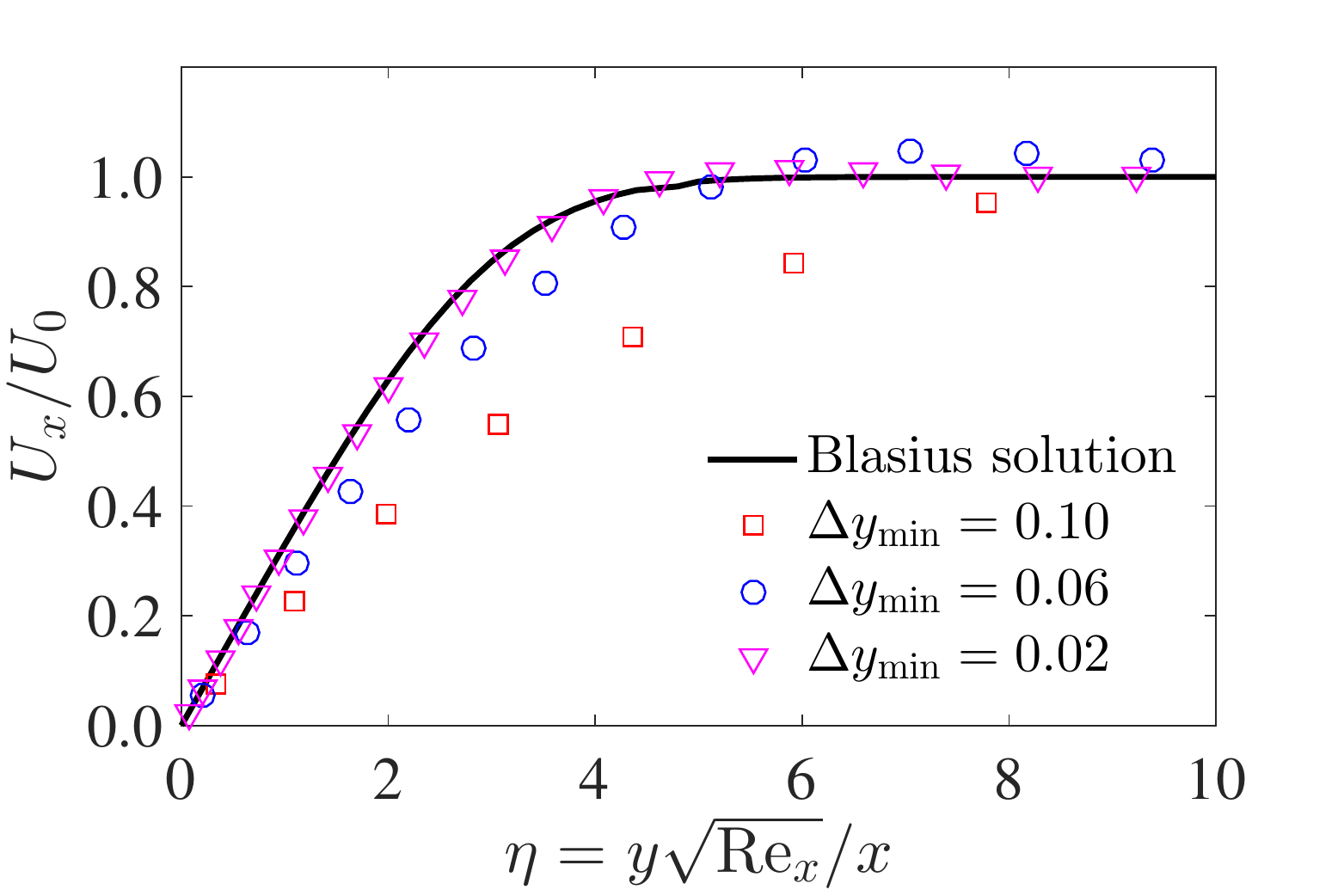}} \\
\subfloat[]{\includegraphics[width=0.48\textwidth]{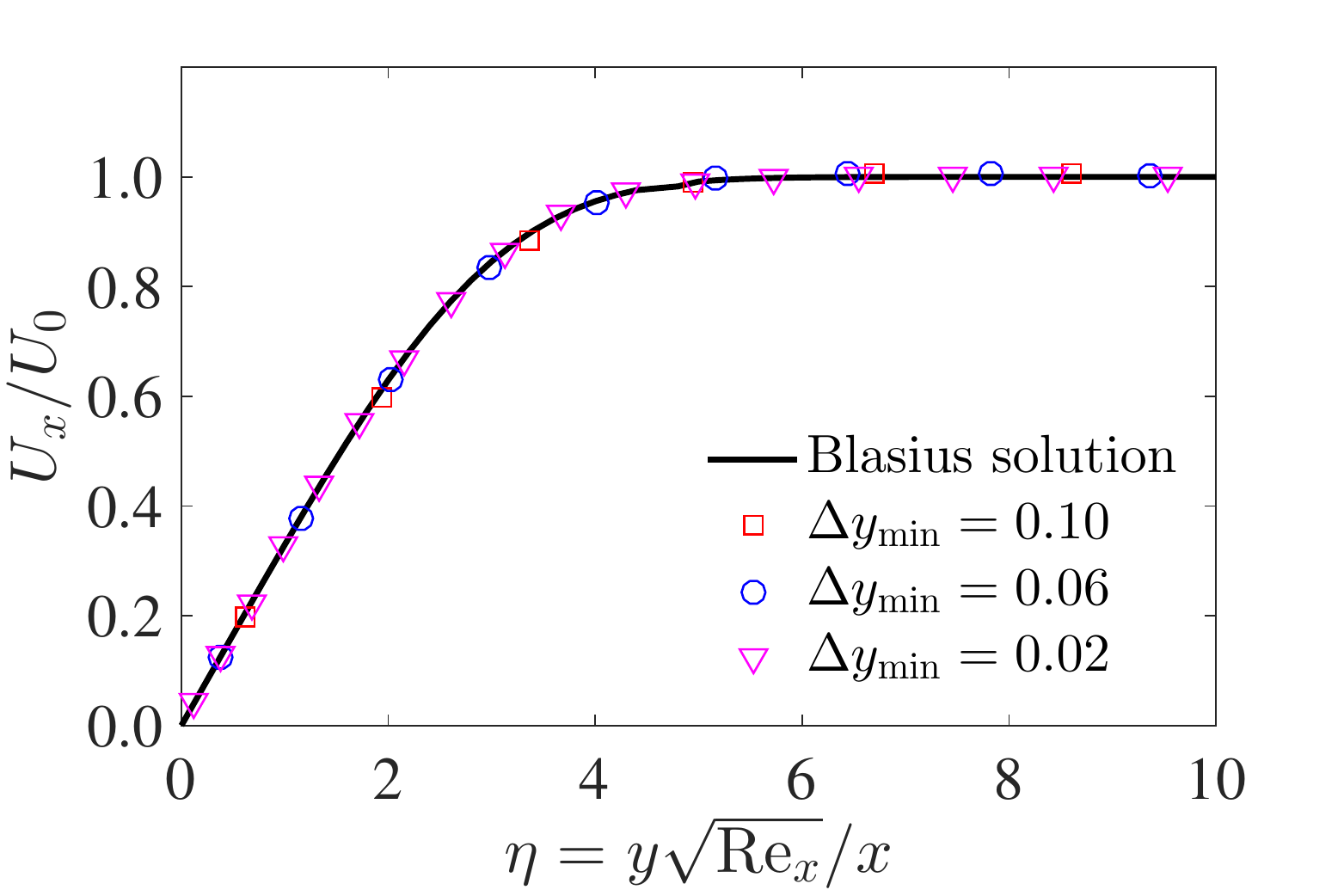}}
\quad
\subfloat[]{\includegraphics[width=0.48\textwidth]{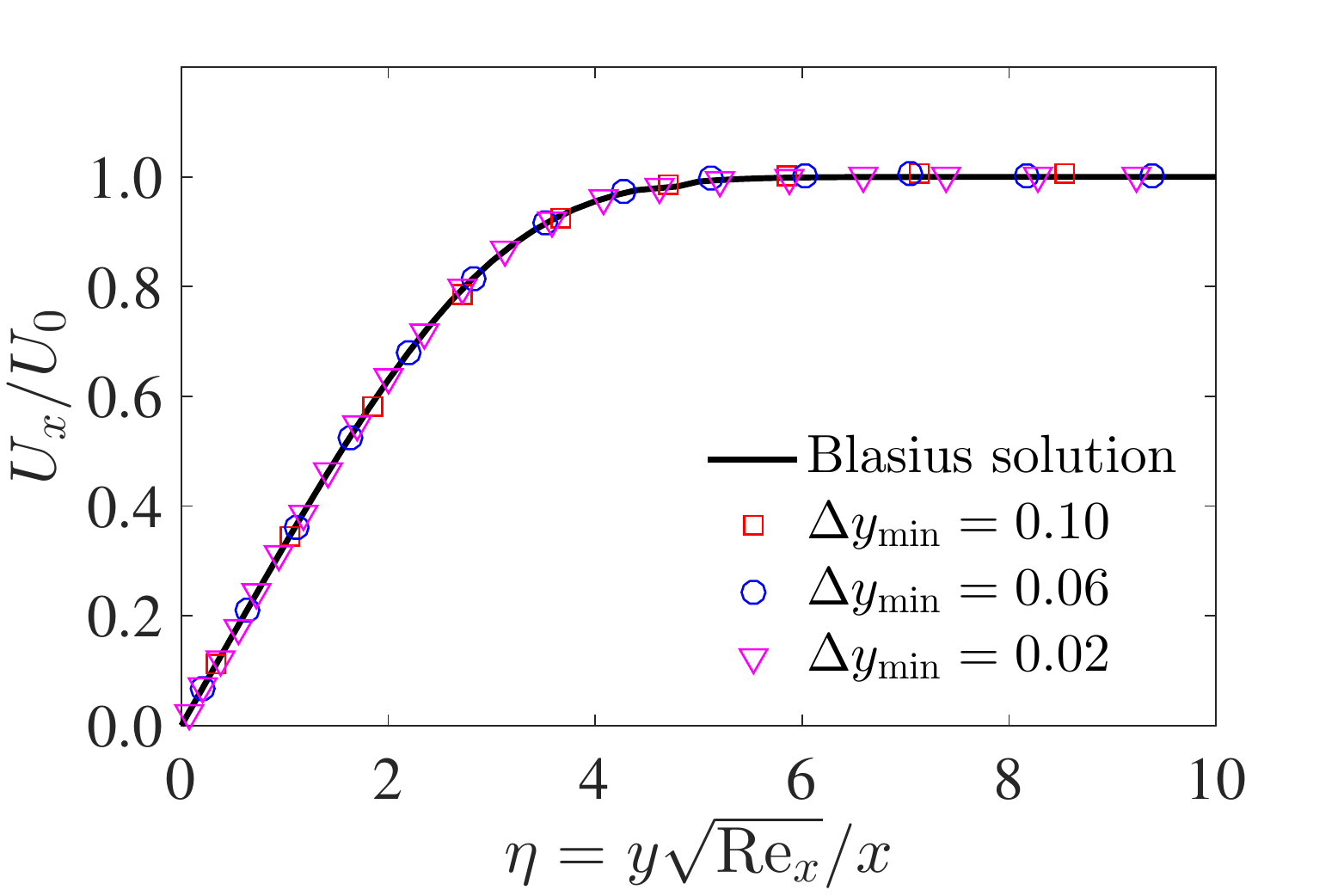}}
\caption{
Horizontal velocity profiles in the boundray layer calculated by the BKG scheme and the DUGKS with different mesh resolutions, CFL=0.5.
 Top: BGK scheme; Bottom: DUGKS; Left: results at x=6.4381; Right: results at x=21.5082.}
\label{fig:bl_u}
\end{figure}

\begin{figure}[htbp]
\centering
\subfloat[]{\includegraphics[width=0.48\textwidth]{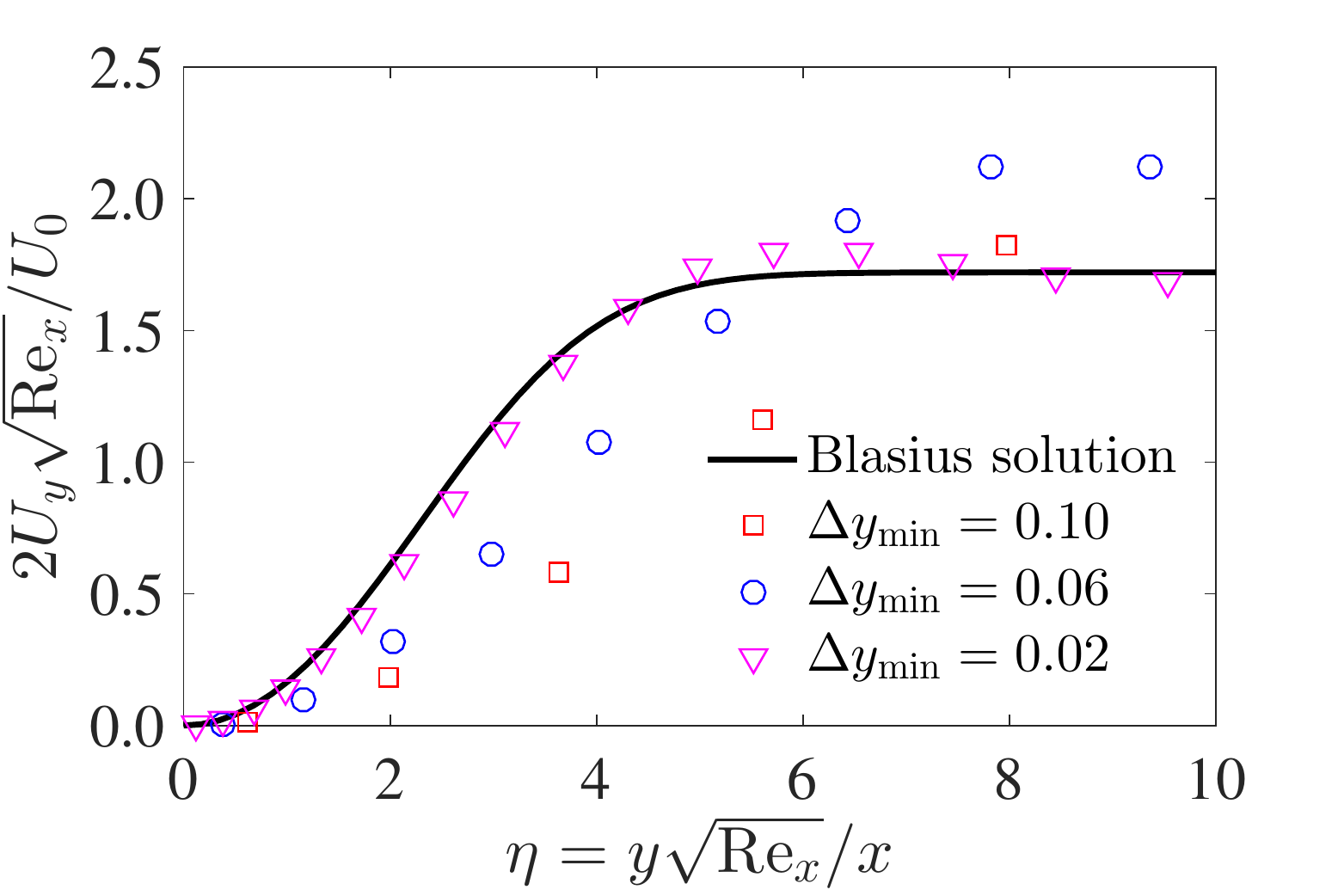}}
\quad
\subfloat[]{\includegraphics[width=0.48\textwidth]{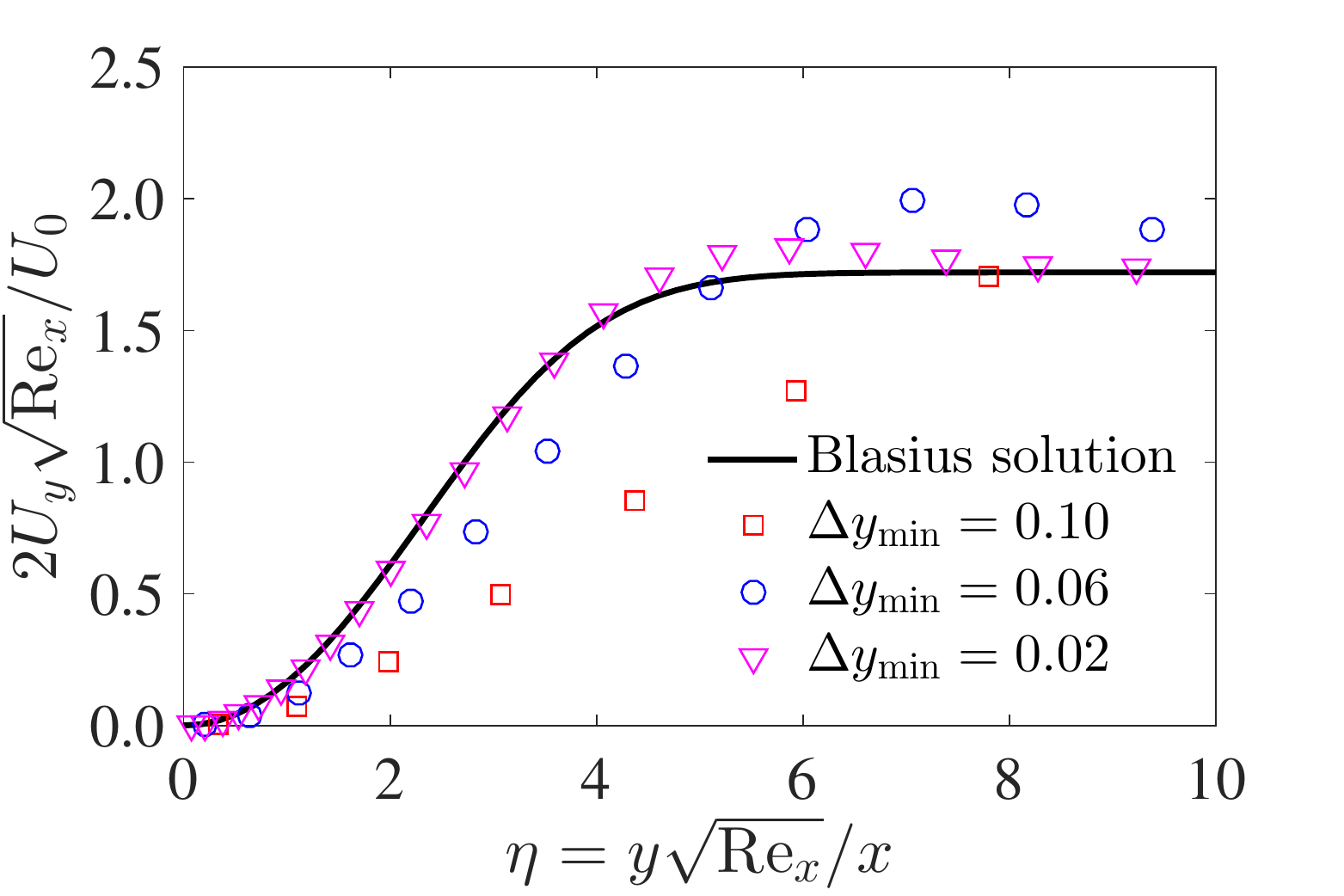}} \\
\subfloat[]{\includegraphics[width=0.48\textwidth]{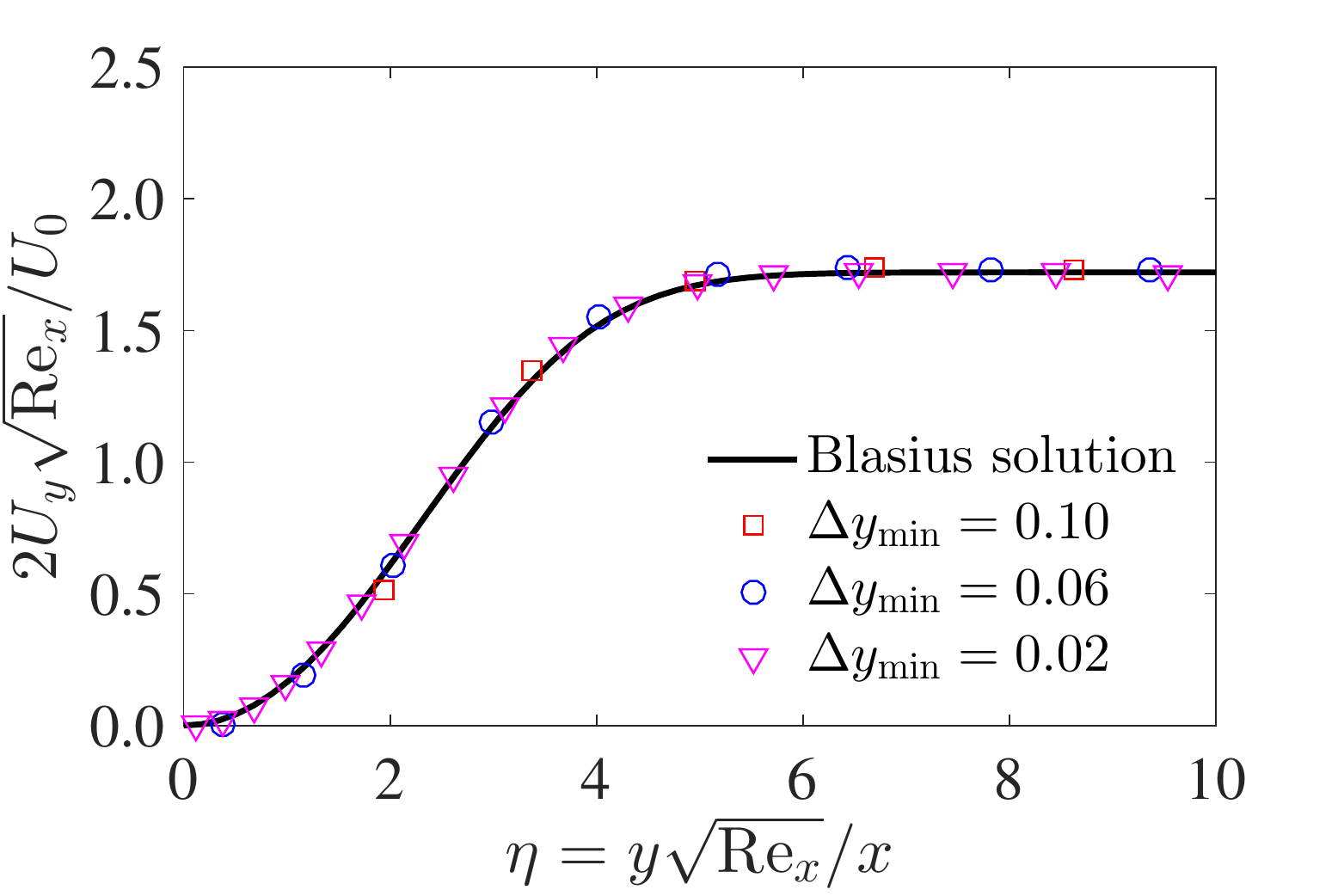}}
\quad
\subfloat[]{\includegraphics[width=0.48\textwidth]{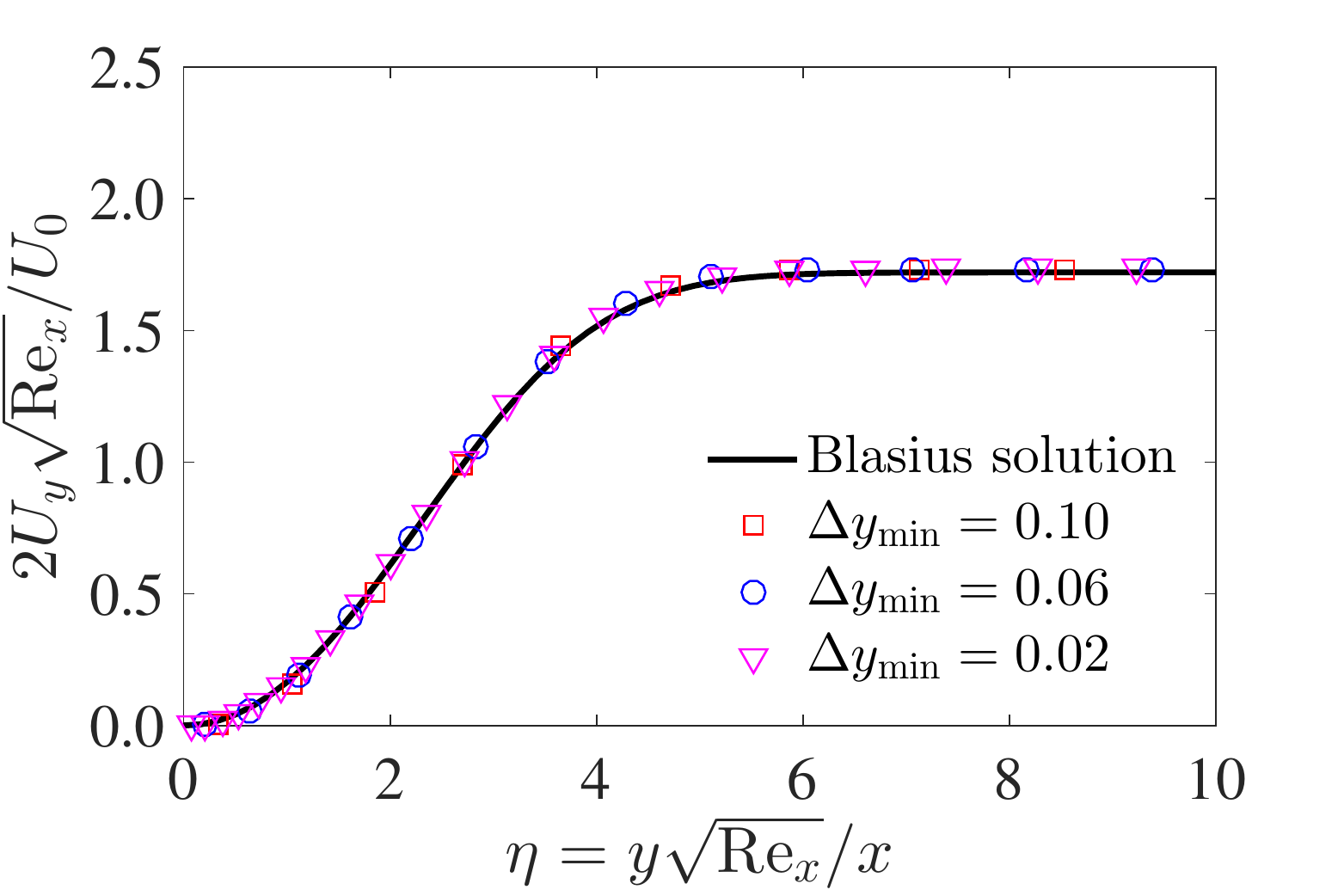}}
\caption{
Vertical velocity profiles in the boundary layer calculated by the BKG scheme and the DUGKS with different mesh resolutions, CFL=0.5.
 Top: BGK scheme; Bottom: DUGKS; Left: results at x=6.4381; Right: results at x=21.5082.
}
\label{fig:bl_v}
\end{figure}

We simulate the flow with different mesh resolutions by adjusting the parameter $\Delta y_{\text{min}}$ from $0.02$ to $0.1$.
The CFL number is fixed at 0.5.
The velocity profiles at $X_1=6.4381$ and $X_2=21.5082$ predicted by the BKG and the DUGKS methods together with the Blasius solutions are shown in Figs.~\ref{fig:bl_u}-\ref{fig:bl_v}.
The horizontal velocity is scaled by $U_0$, and the vertical velocity is scaled by $U_0/2\sqrt{\text{Re}_x}$, where $\text{Re}_x$ is the local Reynolds number defined by $U_0x/\nu$.
\begin{figure}[htbp]
\centering
\subfloat[]{\includegraphics[width=0.48\textwidth]{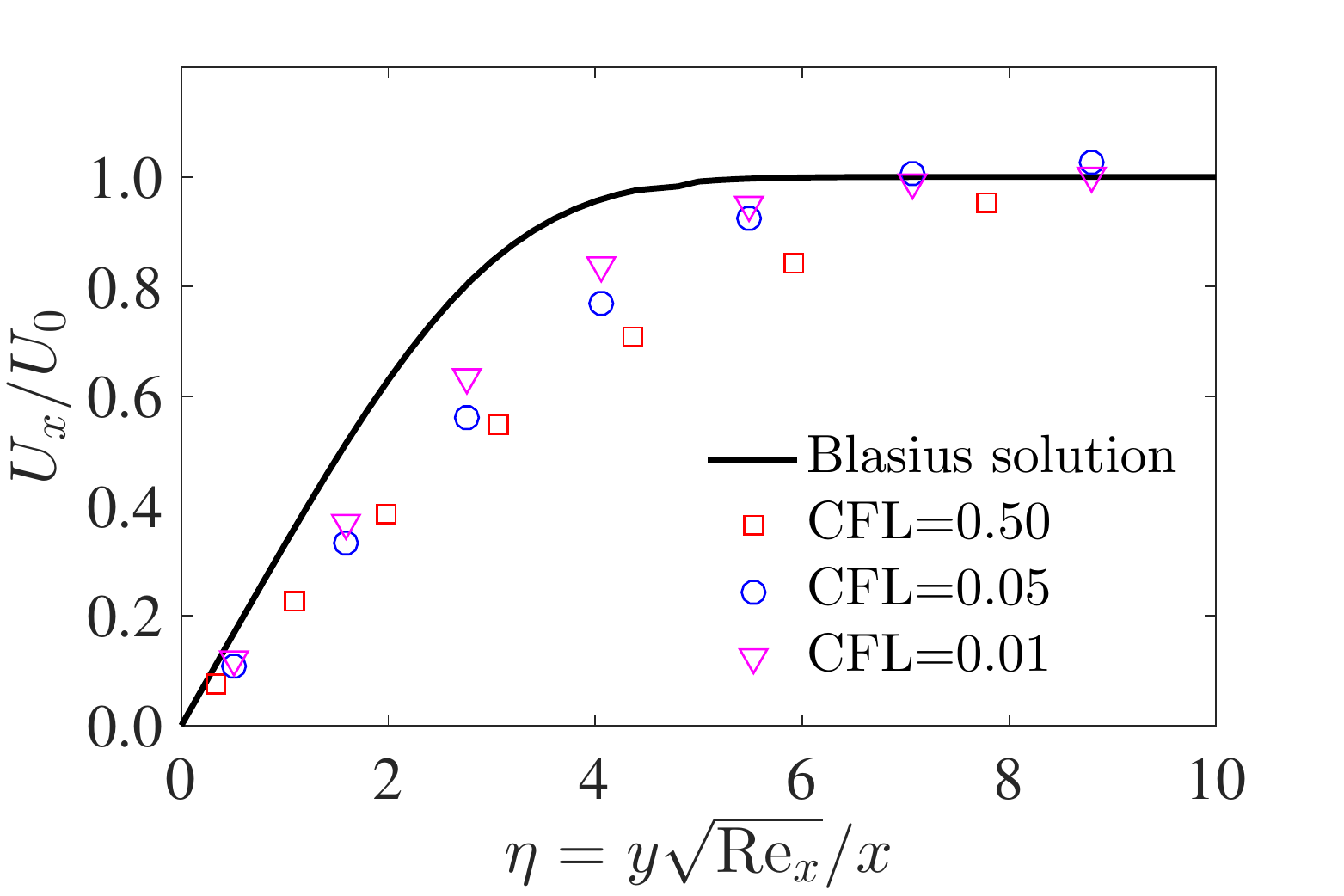}}
\quad
\subfloat[]{\includegraphics[width=0.48\textwidth]{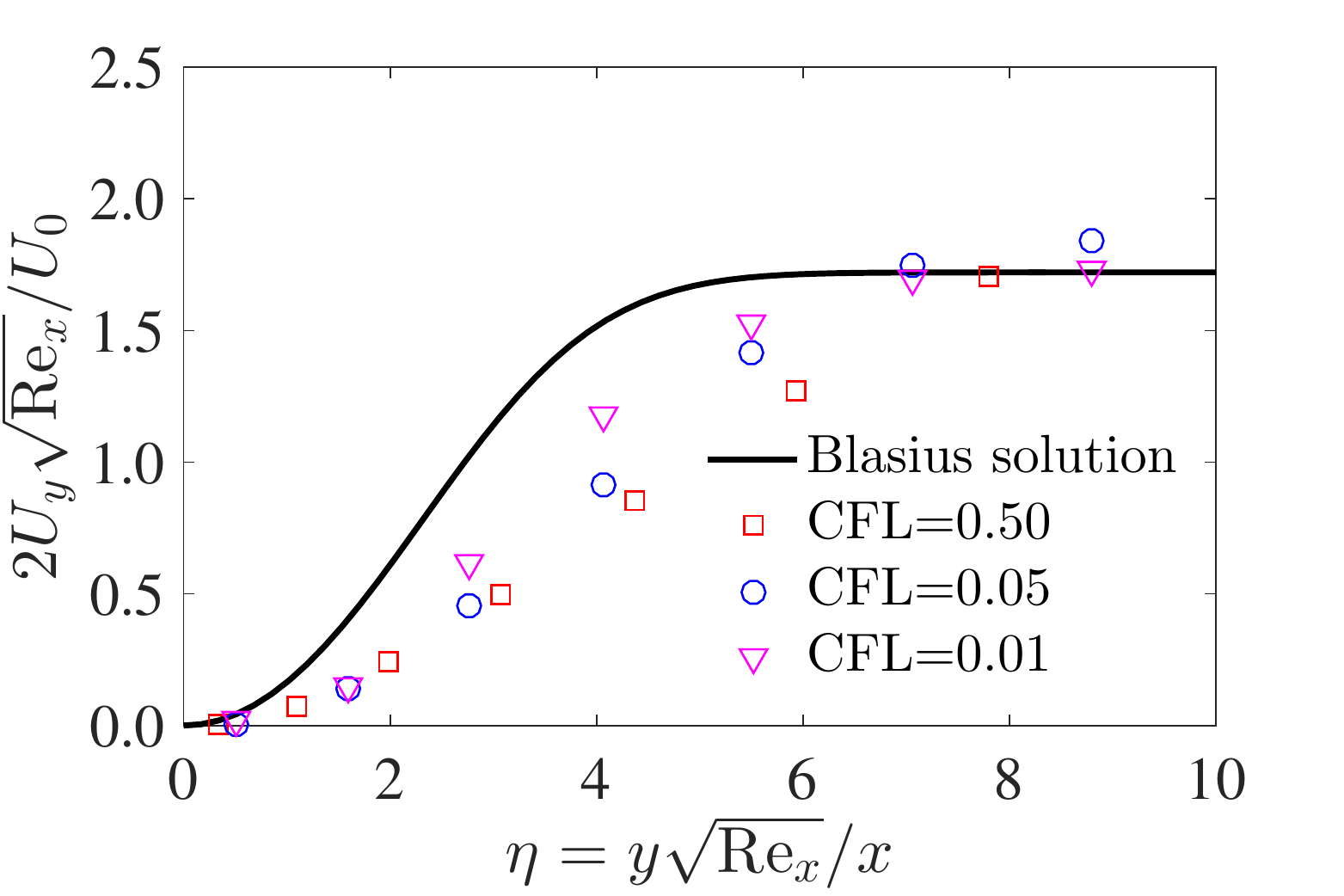}}
\caption{
  Horizontal and vertical velocity profiles at $x=21.5082.$ for boundary layer flow case, calculated by the BKG scheme using $\Delta y_\text{min} = 0.1$ and different CFL numbers.
 (a) Horizontal velocity profile. (b) Vertical velocity profile.
}\label{fig:bl_bkg_cfl}
\end{figure}
From these results, we can observe that,
the boundary layer can be captured accurately by the DUGKS with the three meshes, particularly, with the coarsest mesh ($\Delta y_{\text{min}}=0.1$)
there are only 4 cells in the boundary layer at $x=X_1$.
On the other hand, the BKG scheme can't give satisfactory results even with the finest mesh ($\Delta y_\text{min} = 0.02$),
as shown in Fig.~\ref{fig:bl_u}(a), Fig.~\ref{fig:bl_u}(b), Fig.~\ref{fig:bl_v}(a) and Fig.~\ref{fig:bl_v}(b).
It is also observed that the results of the BKG scheme are quite sensitive to the mesh employed.
These results suggest again that the DUGKS is more robust than the BKG scheme.

Like the cavity flow case, we reduce the time step in the BKG simulation to examine the effects of time step.
The computation is carried out on the mesh of $\Delta y_{\text{min}}=0.1$ and the CFL number varies from $0.5$ to $0.01$.
The velocity profiles are presented in Fig.~\ref{fig:bl_bkg_cfl}.
It can be seen that the use of a small time step can improve the accuracy, but the deviations from the Blasius solution are still obvious even with CFL=0.01.

\section{Conclusions}
In this paper, the performance of two kinetic schemes, i.e., the BKG scheme and the DUGKS is compared.
Both of them can remove the time step restriction which is commonly seen in many off-lattice Boltzmann schemes.
A theoretical analysis in the finite-volume framework demonstrates that the two methods differ only in the constructions of numerical flux.
The BKG scheme treats the collision integral with the one-point quadrature when integrate the BGK equation along the characteristic line to evaluate the numerical flux, while DUGKS computes it with the trapezoidal quadrature.
Consequently, the DUGKS is more accurate and stable than the BKG scheme.

The numerical results of three test cases, including unsteady and steady flows, confirm that the
the DUGKS is more accurate and stable than the BKG scheme on the same computing configurations, especially for high Reynolds number flows.
It is also observed that the DUGKS is stable as long as $\text{CFL}<1$, while the BKG scheme's stability degrades quickly as the CFL number goes beyond 0.5.
We attribute this to the implicit treatment of the collision term of the DUGKS when evaluating the numerical flux.
Furthermore, the results show that the DUGKS is more insensitive to mesh resolutions than the BKG method.
Numerical results also demonstrate that the BKG scheme is about one time faster than the DUGKS on a same mesh.
However, it should be noted that the latter can achieve an accurate solution with a much finer mesh, suggesting that it can be more efficient for flow computations than the BGK scheme.
In summary, the theoretical analysis and numerical results demonstrate that the DUGKS can serve as an efficient method for simulating continuum flows, although it is not limited to such flow regime.

\section*{Acknowledgments}
The authors thank Prof.~Kun Xu for many helpful discussions, and thank Dr.~Weidong Li for providing the benchmark data of the laminar boundary layer.
This study is financially supported by the National Science Foundation of China (Grant No. 51125024)
and the Fundamental Research Funds for the Central Universities (Grant No. 2014TS119).

\end{document}